\documentclass[proceedings]{ESI} 
\usepackage{epsfig}    
\usepackage{clpsref}    

\title{Continuum Strong QCD:\\
Confinement and Dynamical Chiral Symmetry Breaking}

\author{C.D. Roberts\\Physics
 Division, Bldg. 203, Argonne National Laboratory, Argonne IL 60439-4843,
 USA.}

\conference{Confinement @ ESI, May-July 2000}

\abstract{Continuum strong QCD is the application of models and continuum
quantum field theory to the study of phenomena in hadronic physics, which
includes; e.g., the spectrum of QCD bound states and their interactions.
Herein I provide a Dyson-Schwinger equation perspective, focusing on
qualitative aspects of confinement and dynamical chiral symmetry breaking in
cold, sparse QCD, and also elucidating consequences of the axial-vector
Ward-Takahashi identity and features of the heavy-quark limit.}

\keywords{Bethe-Salpeter Equation, Confinement, Dynamical Chiral Symmetry
Breaking, Dyson-Schwinger Equations, Hadron Physics, Heavy-quark Physics}

\newcommand{\lsim}{\mathrel{\rlap{\lower4pt\hbox{\hskip0pt$\sim$}}
\raise1pt\hbox{$<$}}}
\newcommand{\gsim}{\mathrel{\rlap{\lower4pt\hbox{\hskip0pt$\sim$}}
\raise1pt\hbox{$>$}}}           
\newcommand{\sfrac}[2]{\mbox{\footnotesize $\frac{#1}{#2}$}}

\renewcommand{\theenumii}{\arabic{enumi}.\arabic{enumii}}

\begin{document} 

\hyphenation{ap-pro-xi-ma-ted}
\hyphenation{mo-men-tum} 

\section*{Table of Contents}
\begin{enumerate}
\item Introduction \dotfill \pageref{sec.intro}
\item Quark Propagator \dotfill \pageref{sec.quark}
\begin{enumerate}
\item {\small\it Gluon Propagator} \dotfill \pageref{sub.gluonpropagator}
\item {\small\it Quark-gluon Vertex} \dotfill \pageref{sub.quarkgluonvertex}
\end{enumerate}
\item Dynamical Chiral Symmetry Breaking \dotfill \pageref{sec.dcsb}
\item Confinement \dotfill \pageref{sec.conf}
\begin{enumerate}
\item {\small\it Three-dimensional QED} \dotfill \pageref{sub.qed3}
\end{enumerate}
\item Gap Equation's Kernel \dotfill \pageref{sec.kernel}
\begin{enumerate}
\item {\small\it Critical Interaction Tension} \dotfill
\pageref{sub.polarisationtension}
\end{enumerate}
\item Gluon DSE \dotfill \pageref{sec.gluon}
\item Bethe-Salpeter Equation \dotfill \pageref{sec.bound}
\begin{enumerate}
\item {\small\it Heavy-quark Limit} \dotfill \pageref{sub.heavy}
\end{enumerate}
\item Epilogue \dotfill \pageref{sec.contemporary}
\item References \dotfill \pageref{References}
\end{enumerate}

\section{Introduction}
\label{sec.intro}
A primary goal in continuum strong QCD is to develop an intuitive
understanding of the spectrum and interactions of hadrons in terms
of\linebreak QCD's elementary degrees of freedom.  In addressing this an
efficacious strategy is to employ a single framework in calculating a
prodigious number of observables.  With the aim being an elucidation of
hadronic structure and nonperturbative aspects of QCD, a focus on the
electroweak interactions of hadrons is most useful because in this case the
probe is well understood and the features of the hadronic target are
unambiguously under scrutiny.  Both confinement and dynamical chiral symmetry
breaking (DCSB) play key roles in determining these features, and are the
main themes of this discourse.

As reviewed in Ref.~\cite{bastirev}, the last decade has seen a modest
renaissance in the use of Dyson-Schwinger equations (DSEs)~\cite{cdragw}: in
exploring\linebreak their formal foundation and in their phenomenological
application.  They provide the medium for this discussion.

The DSEs are a nonperturbative means of analysing a quantum field theory.
Derived from a theory's Euclidean space generating functional, they are an
enumerable infinity of coupled integral equations whose solutions are the
$n$-point Schwinger functions [Euclidean Green functions], which are the same
matrix elements estimated in numerical simulations of lattice-QCD.  In
theories with elementary fer\-mi\-ons, the simplest of the DSEs is the {\it
gap} equation, which is basic to studying dynamical symmetry breaking in
systems as disparate as ferromagnets, superconductors and QCD.  The gap
equation is a good example because it is familiar and has all the properties
that characterise each DSE: its solution is a $2$-point function [the fermion
propagator] while its kernel involves higher $n$-point functions; e.g., in a
gauge theory, the kernel is constructed from the gauge-boson $2$-point
function and fermion--gauge-boson vertex, a $3$-point function; a
weak-coupling expansion yields all the diagrams of perturbation theory; and
solved self-consistently, the solution of the gap equation exhibits
nonperturbative effects unobtainable at any finite order in perturbation
theory; e.g, dynamical symmetry breaking.

The coupling between equations; i.e., the fact that the equation for a given
$m$-point function always involves at least one $n>m$-point function,
necessitates a truncation of the tower of DSEs in order to define a tractable
problem.  One systematic and familiar truncation is a weak coupling expansion
to reproduce perturbation theory.  However, that precludes the study of
nonperturbative phenomena and hence something\linebreak else is needed for
the investigation of strongly interacting systems and bound state phenomena.

In analysing the ferromagnetic transition, the Hartree-Fock approximation
yields qualitatively reliable information and in QCD its analogue: the
rainbow truncation, has proven efficacious.  However, {\it a priori} it can
be difficult to judge whether a given truncation will yield reliable results
and a systematic improvement is not always obvious.  It is here that some
model-dependence enters but that is not new, being typical in the study of
strongly-interacting few- and many-body systems.  

To proceed with the DSEs one just employs a truncation and explores its
consequences, applying it to different systems and constraining it, where
possible, by comparisons with experimental data, and also with other
theoretical approaches on their common domain of application.  In this way a
reliable truncation can be identified, and then attention paid to
understanding the keystone of its success and improving its foundation.  This
pragmatic approach has proven rewarding in strong QCD, not least because a
correctly-executed weak coupling expansion is\linebreak guaranteed to match
onto perturbation theory so that modelling is restricted to the infrared
domain.

\section{Quark Propagator}
\label{sec.quark}
Obviously it is only possible to study DCSB in theories with a well-defined
chiral limit.  Asymptotically free theories such as QCD are in this class.  A
useful starting point for any discussion of DCSB is the renormalised
quark-DSE [see App.~\ref{appEM} for the Euclidean metric conventions used
herein]:
\begin{eqnarray}
\label{gendse}
\lefteqn{S(p)^{-1}  =  Z_2 \,(i\gamma\cdot p + m_{\rm bare})}\\
&& \nonumber +\, Z_1 \int^\Lambda_q \, g^2 D_{\mu\nu}(p-q)
\frac{\lambda^a}{2}\gamma_\mu S(q) \Gamma^a_\nu(q,p) \,,
\end{eqnarray}
where $D_{\mu\nu}(k)$ is the renormalised dressed-gluon propagator,
$\Gamma^a_\nu(q;p)$ is the renormalised dressed-quark-gluon vertex, $m_{\rm
bare}$ is the $\Lambda$-dependent\linebreak current-quark bare mass that
appears in the Lagrangian and $\int^\Lambda_q := \int^\Lambda d^4 q/(2\pi)^4$
represents\linebreak mnemonically a {\em translationally-invariant}
regularisation of the integral, with $\Lambda$ the regularisation mass-scale.
The final step in any calculation is to remove the regularisation by taking
the limit $\Lambda \to \infty$.

Using a translationally invariant regularisation makes possible the
preservation of Ward-Takahashi identities, which is crucial; e.g.,
in\linebreak studying DCSB \cite{truncscheme}.  One implementation
well-suited to a nonperturbative solution of the DSE is Pauli-Villars
regularisation, in which quarks interact with an additional massive
gluon-like\linebreak vector boson: \mbox{mass$\,\sim \Lambda$}, that
decouples as $\Lambda\to \infty$~\cite{mr97}.  An alternative is a numerical
implementation of dimensional regularisation, which, although more
cumbersome, can provide the necessary check of
scheme-in\-de\-pen\-dence~\cite{sizer}.

In Eq.~(\ref{gendse}), $Z_1(\zeta^2,\Lambda^2)$ and $Z_2(\zeta^2,\Lambda^2)$
are the quark-gluon-vertex and quark wave function renormalisation constants,
which depend on the renormalisation point, $\zeta$, and the regularisation
mass-scale, as does the mass renormalisation constant
\begin{equation}
\label{Zmass}
Z_m(\zeta^2,\Lambda^2) = Z_4(\zeta^2,\Lambda^2)/Z_2(\zeta^2,\Lambda^2) ,
\end{equation}
with the renormalised mass given by
\begin{equation}
\label{mzeta}
m(\zeta) := m_{\rm bare}(\Lambda)/Z_m(\zeta^2,\Lambda^2).
\end{equation}
Although I have suppressed the flavour label, $S$, $\Gamma^a_\mu$ and $m_{\rm
bare}$ depend on it.  However, one can always use a flavour-independent
renormalisation scheme, which I assume herein, and hence all the
renormalisation constants are flavour-in\-de\-pen\-dent~\cite{mr97}.

The solution of Eq.~(\ref{gendse}) has the form
\begin{eqnarray}
\label{sinvp}
S(p) &  = & \frac{1}
{i \gamma\cdot p \,A(p^2,\zeta^2) + B(p^2,\zeta^2)}\,,\\
        &  = & \frac{Z(p^2,\zeta^2)}
                {i\gamma\cdot p + M(p^2,\zeta^2)}\,.
\end{eqnarray}
The functions $A(p^2,\zeta^2)$, $B(p^2,\zeta^2)$ embody all the effects of
vector and scalar quark-dressing induced by the quark's interaction with its
own gluon field.  The ratio: $M(p^2,\zeta^2)$, is the quark mass function and
a {\it pole} mass; i.e., the on-shell mass, would be the solution of
\begin{equation}
m_{\rm pole}^2 - M^2(p^2=-m_{\rm pole}^2,\zeta^2) = 0.
\end{equation}
A widely posed conjecture is that confinement rules out a solution of this
equation~\cite{gastao}, and that is discussed further in Sec.~\ref{sec.conf}.

Equation~(\ref{gendse}) must be solved subject to a renormalisation
[boundary] condition, and\linebreak because the theory is asymptotically free
it is practical and useful to impose the requirement that at a large
spacelike $\zeta^2$
\begin{equation}
\label{renormS}
\left.S(p)^{-1}\right|_{p^2=\zeta^2} = i\gamma\cdot p + m(\zeta)\,,
\end{equation}
where $m(\zeta)$ is the renormalised current-quark\linebreak mass at the
scale $\zeta$.  By ``large'' here I mean $\zeta^2 \gg \Lambda_{\rm QCD}^2$ so
that in quantitative, model studies extensive use can be made of matching
with the results of perturbation theory.  It is the ultraviolet stability of
QCD; i.e., the fact that perturbation theory is valid at large spacelike
momenta, that makes possible a straightforward definition of the chiral
limit.  It also provides the starkest contrast to strong-coupling QED, whose
rigorous definition remains an instructive challenge \cite{cdragw,manuel}.

Multiplicative renormalisability in gauge theories entails that
\begin{equation}
\frac{A(p^2,\zeta^2)}{A(p^2,\tilde\zeta^2)\rule{0ex}{2ex}}
= \frac{Z_2(\zeta^2,\Lambda^2)}{Z_2(\tilde\zeta^2,\Lambda^2)\rule{0ex}{2ex}}
= A(\tilde\zeta^2,\zeta^2) 
= \frac{1}{A(\zeta^2,\tilde\zeta^2)\rule{0ex}{2ex}}
\end{equation}
and beginning with Ref.~\cite{cp90} this relation has been used efficaciously
to build realistic {\it Ans\"atze} for the fermion--photon vertex in quenched
QED.  A systematic approach to such nonperturbative improvements is
developing~\cite{dongbashir} and these improvements continue to provide
intuitive guidance in QED, where they complement the perturbative calculation
of the vertex~\cite{ayse}.  They are also useful in exploring model
dependence in QCD studies \cite{fredthesis}.

At one loop in QCD perturbation theory
\begin{equation}
\label{z2mu}
Z_2(\zeta^2,\Lambda^2) 
= \left[\frac{ \alpha(\Lambda^2) }{\alpha(\zeta^2)}
        \right]^{-\gamma_F/\beta_1}\!,
\end{equation}
$\gamma_F= \sfrac{2}{3}\xi,\; \beta_1= \sfrac{1}{3}N_f - \sfrac{11}{2},$ and
at this order the running strong-coupling is
\begin{equation}
\label{alphaq2}
\alpha(\zeta^2) = \frac{\pi} {\rule{0mm}{1.2\baselineskip} -\sfrac{1}{2}
             \beta_1 \ln\left[\zeta^2/\Lambda_{\rm QCD}^2\right]}\,.
\end{equation}
In Landau gauge: $\xi = 0$, so $Z_2 \equiv 1$ at one loop order.  This, plus
the fact that Landau gauge is a fixed point of the renormalisation
group\linebreak\ [Eq.~(\ref{landau})], makes it the most useful covariant
gauge for model studies.  It also underlies the quantitative accuracy of
Landau gauge rainbow truncation estimates of the critical coupling in strong
QED~\cite{mrpAdelaideadnan}.  In a self consistent solution of
Eq.~(\ref{gendse}), $Z_2 \neq 1$ even in Landau gauge but, at large
$\zeta^2$, the $\zeta$-dependence is very weak.  However, as will become
evident, in studies of realistic QCD models this dependence becomes
significant for $\zeta^2 \lsim 1$--$2\,$GeV$^2$, and is driven by the same
effect that causes DCSB.

The dressed-quark mass function: $M(p^2,\zeta^2)$ $=$
$B(p^2,\zeta^2)/A(p^2,\zeta^2)$, is independent of the renormalisation point;
i.e., with $\zeta\neq \tilde\zeta$
\begin{equation}
\label{Mrpi}
M(p^2,\zeta^2) = M(p^2,\tilde\zeta^2):= M(p^2)\,, \; \forall\, p^2.
\end{equation}
It is a function only of $p^2/\Lambda_{\rm QCD}^2$, which is another
constraint on models.  At one loop order the running [or renormalised] mass
\begin{equation}
\label{masanom}
m(\zeta) = M(\zeta^2) = \frac{\hat m} {\left(\rule{0mm}{1.2\baselineskip}
\sfrac{1}{2}\ln\left[\zeta^2/\Lambda_{\rm QCD}^2
\right]\right)^{\gamma_m}}\,,
\end{equation}
$\gamma_m= 12/(33-2 N_f)$, where $\hat m$ is the renormalisation point
independent current-quark mass, and the mass renormalisation constant is,
Eq.~(\ref{Zmass}),
\begin{equation}
\label{zmdef}
Z_m(\zeta^2,\Lambda^2) 
= \left[\frac{ \alpha(\Lambda^2) }{\alpha(\zeta^2)}\right]^{\gamma_m}\,.
\end{equation}
The mass anomalous dimension, $\gamma_m$, is independent of the gauge
parameter to all orders in perturbation theory and for two different quark
fla\-vours the ratio: 
\begin{equation}
m_{f_1}(\zeta) / m_{f_2}(\zeta) = \hat m_{f_1}/\hat
m_{f_2}, 
\end{equation}
which is independent of the renormalisation point and of the renormalisation
scheme.  The chiral limit is unambiguously defined by
\begin{equation}
\label{chirallimit}
{\bf chiral~limit}:\; \hat m = 0\,.
\end{equation}

I reiterate now that a weak coupling expansion of Eq.~(\ref{gendse}) yields
each of the diagrams in perturbation theory that contributes to the quark
self energy.  However, every one of those contributions to $B(p^2,\zeta^2)$
is proportional to $\hat m$ and therefore vanishes in the chiral limit; i.e.,
$B(p^2,\zeta^2)\equiv 0$ at every order in perturbation theory.  

One finds, in fact, that in the chiral limit there is no scalar mass-like
divergence in the calculation of the self energy.  This is manifest in the
quark DSE, with Eq.~(\ref{gendse}) capable of yielding, in addition to the
perturbative result: $B(p^2,\zeta^2)\equiv 0$, a solution
$M(p^2)=B(p^2,\zeta^2)/A(p^2,\zeta^2)\neq 0$ that is power-law suppressed in
the ultraviolet: $M(p^2) \sim 1/p^2$, guaranteeing {\it convergence} of the
associated integral {\it without subtraction}\label{`without subtraction'}.
This is dynamical chiral symmetry breaking
\begin{equation}
{\bf DCSB:}\;M(p^2)\neq 0\,\; {\rm when} \;\hat m = 0\,.
\end{equation}
As we shall see, in QCD this is possible if and only if the quark condensate
is nonzero: the criteria are equivalent, and its existence places constraints
on the kernel in Eq.~(\ref{gendse}), as discussed further in
Sec.~\ref{sec.kernel}.

\subsection{Gluon Propagator}
\label{sub.gluonpropagator}
That kernel is constructed from the dressed-gluon propagator and the
dressed-quark-gluon\linebreak vertex, and encodes in Eq.~(\ref{gendse}) all
effects of the quark-quark interaction.  In a covariant gauge the
renormalised dressed-gluon propagator is
\begin{equation}
\label{gluoncovariant}
D_{\mu\nu}(k) = \left( \delta_{\mu\nu} - \frac{k_\mu k_\nu}{k^2}\right)
                \frac{d(k^2,\zeta^2)}{k^2} + \xi\,\frac{k_\mu k_\nu}{k^4}\,,
\end{equation}
where $d(k^2,\zeta^2) = 1/[1+\Pi(k^2,\zeta^2)]$, with $\Pi(k^2,\zeta^2)$ the
renormalised gluon vacuum polarisation for which the conventional
renormalisation condition is 
\begin{equation}
\label{renormPi}
\Pi(\zeta^2,\zeta^2) = 0\,;\; {\rm i.e.,}\; d(\zeta^2,\zeta^2) =1\,.
\end{equation}

For the dressed-gluon propagator, multiplicative renormalisability entails
\begin{equation}
\label{mrPi}
\frac{d(k^2,\zeta^2)}{d(k^2,\tilde\zeta^2)\rule{0ex}{2ex}}
= \frac{Z_3(\tilde\zeta^2,\Lambda^2)}{Z_3(\zeta^2,\Lambda^2)\rule{0ex}{2ex}}
= d(\zeta^2,\tilde\zeta^2)
= \frac{1}{d(\tilde\zeta^2,\zeta^2)\rule{0ex}{2ex}}\,,
\end{equation}
and at one loop in perturbation theory
\begin{equation}
\label{z3mu}
Z_3(\zeta^2,\Lambda^2) 
= \left[\frac{ \alpha(\Lambda^2) }{\alpha(\zeta^2)}
        \right]^{-\gamma_1/\beta_1},
\end{equation}
$\gamma_1 = \sfrac{1}{3}N_f - \sfrac{1}{4}(13 - 3\,\xi)$.  The gauge
parameter is also renormalisation point dependent; i.e., the renormalised
theory has a running gauge parameter.  However, because of Becchi-Rouet-Stora
[BRST or gauge] invariance, there is no new dynamical information in that:
its evolution is completely determined by the gluon wave function
renormalisation constant
\begin{equation}
\xi(\zeta^2) = Z_3^{-1}(\zeta^2,\Lambda^2)\,\xi_{\rm bare}(\Lambda)\,.
\end{equation}
One can express $\xi(\zeta^2)$ in terms of a renormalisation point invariant
gauge parameter: $\hat \xi$, which is an overall multiplicative factor in the
formula and hence
\begin{equation}
\label{landau}
{\bf Landau~Gauge:}\; \hat \xi = 0 \Rightarrow \xi(\zeta^2)\equiv 0
\end{equation}
at all orders in perturbation theory; i.e., Landau gauge is a fixed point of
the renormalisation group.

\subsection{Quark-gluon Vertex}
\label{sub.quarkgluonvertex}
The other element of the kernel, the renormalised dressed-quark-gluon vertex,
has the form
\begin{equation}
\Gamma^a_\nu(k,p) = \frac{\lambda^a}{2} \Gamma_\nu(k,p)\,;
\end{equation}
i.e., the colour matrix structure factorises.  It is a fully amputated
vertex, which means all the analytic structure associated with elementary
excitations has been eliminated.  To discuss this further I introduce the
notion of a particle-like singularity.

\hspace*{-\parindent}\underline{\em A Particle-like Singularity}~ is one of
the form: $P= k-p$,
\begin{equation}
\label{plsing}
\frac{1}{(P^2+b^2)^{\alpha}},\; \alpha\in(0,1]\,.
\end{equation}
If the vertex possesses such a singularity then its $P$-dependence can be
expressed via a non-negative spectral density, which is impossible if
$\alpha>1$.  $\alpha=1$ is the ideal case of an isolated $\delta$-function
distribution in the spectral densities and hence an isolated free-particle
pole.  $\alpha\in(0,1)$ corresponds to an accumulation at the particle pole
of branch points associated with multiparticle production, as occurs with the
electron propagator in QED because of photon dressing.

The dressed-quark-gluon vertex is a {\it fully amputated} $3$-point function.
Therefore in this case the presence of such a singularity would entail the
existence of a flavour singlet composite [quark-antiquark bound state] with
colour octet quantum numbers and mass $m=b$.  [The bound state amplitude
follows immediately from the associated homogeneous Bethe-Salpeter equation
(BSE), which the singularity allows one to derive.]  However, an excitation
like this must not exist as an asymptotic state: that would violate the
observational evidence of confinement.  Hence, as discussed further on page
\pageref{discuss.vertex}, I conclude that the vertex should not exhibit a
particle-like singularity, and any modelling of $\Gamma_\mu^a(k,p)$ ought to
be consistent with this constraint.  [NB.\ $\alpha >1$ yields an admissible
non-particle-like singularity.]

Expressing the Dirac structure of $\Gamma_\nu(k,p)$ requires twelve
independent scalar functions:
\begin{equation}
\Gamma_\nu(k,p) = \gamma_\nu \,F_1(k,p,\zeta) + \ldots\,,
\end{equation}
which it is not necessary to articulate fully herein.  A pedagogical
discussion of the perturbative calculation of $\Gamma_\nu(k,p)$ can be found
in Ref.~\cite{pt84} while Refs.~\cite{bc80,vertex} explore its
nonperturbative structure and properties.  I only make $F_1(k,p,\zeta)$
explicit because the renormalisability of QCD entails that it alone is
ultraviolet divergent.  Defining
\begin{equation}
f_1(k^2,\zeta^2):= F_1(k,-k,\zeta)\,,
\end{equation}
the conventional renormalisation boundary condition is
\begin{equation}
f_1(\zeta^2,\zeta^2) = 1\,,
\end{equation}
which is practical because QCD is asymptotically free.  Multiplicative
renormalisability entails
\begin{eqnarray}
\lefteqn{\frac{f_1(k^2,\zeta^2)}{f_1(k^2,\tilde\zeta^2)\rule{0ex}{2ex}}}\\
&& \nonumber
= \frac{Z_1(\zeta^2,\Lambda^2)}{Z_1(\tilde\zeta^2,\Lambda^2)\rule{0ex}{2ex}}
= f_1(\tilde\zeta^2,\zeta^2)
= \frac{1}{f_1(\zeta^2,\tilde\zeta^2)\rule{0ex}{2ex}}\,,
\end{eqnarray}
and at one loop order
\begin{equation}
\label{z1mu}
Z_1(\mu^2,\Lambda^2) = \left[
                \frac{\alpha(\Lambda^2)}{\alpha(\mu^2)}
                        \right]^{-\gamma_\Gamma/\beta_1}\,,
\end{equation}
$\gamma_\Gamma = \sfrac{1}{2}[ \sfrac{3}{4} (3+\xi) + \sfrac{4}{3}\xi]$.

\section{DCSB}
\label{sec.dcsb}
At this point each element in the quark DSE, Eq.~(\ref{gendse}), is defined,
with some of their perturbative properties elucidated, and the question is:
``How does that provide an understanding of DCSB?''  It is best answered
using an example, in which the model-independent aspects are made clear.

The quark DSE is an integral equation and hence its elements must be known at
all values of their momentum arguments, not just in the perturbative domain
but also in the infrared.  While the gluon propagator and quark-gluon vertex
each satisfy their own DSE, that couples the quark DSE to other members of
the tower of equations and hinders rather than helps in solving the gap
equation.  Therefore, as with all applications of the gap equation, one
employs {\it An\-s\"at\-ze} for the interaction elements [$D_{\mu\nu}(k)$ and
$\Gamma_\nu(k,p)$], constrained as much and on as large a domain as possible.
This approach has a long history in exploring QCD \cite{cdragw} and I
illustrate it using the model of Ref.~\cite{mr97}.

The renormalised dressed-ladder truncation of the quark-antiquark scattering
kernel [$4$-point function] is
\begin{eqnarray}
\lefteqn{\bar K(p,q;P)_{tu}^{rs} = g^2(\zeta^2)\, D_{\mu\nu}(p-q) }\\
&& \nonumber 
\left[\rule{0mm}{0.7\baselineskip} \Gamma^a_\mu(p_+,q_+)\,S(q_+) \right]_{tr} 
\,
\left[ \rule{0mm}{0.7\baselineskip}S(q_-)\,\Gamma^a_\nu(q_-,p_-) \right]_{su},
\end{eqnarray}
where $p_\pm=p \,\pm \,P/2$, $q_\pm=q \,\pm \,P/2$, with $P$ the total
momentum of the quark-antiquark pair, and although I use it now I have
suppressed the $\zeta$-dependence of the Schwinger functions.  From
Eqs.~(\ref{renormPi}-\ref{z3mu}) it follows that for $Q^2:=(p-q)^2$ large and
spacelike
\begin{eqnarray}
d(Q^2,\zeta^2)&= &\frac{Z_3(\zeta^2,\Lambda^2)}{Z_3(Q^2,\Lambda^2)}
\,d(\zeta^2,\zeta^2) \\
& = & \left[\frac{\alpha(Q^2)}{\alpha(\zeta^2)}\right]^{\gamma_1/\beta_1}
\end{eqnarray}
\begin{equation}
\Rightarrow  D_{\mu\nu}(p-q) = 
\left[\frac{\alpha(Q^2)}{\alpha(\zeta^2)}\right]^{\gamma_1/\beta_1}\,
\!\!\!D_{\mu\nu}^{\rm free}(p-q)\,.
\end{equation}
Using this and analogous results for the other Schwinger functions then on
the kinematic domain for which $Q^2 \sim p^2\sim q^2$ is large and spacelike
[$g^2(\zeta^2):= 4\pi\alpha(\zeta^2)$]
\begin{eqnarray}
\label{aqquv}
\lefteqn{\bar K(p,q;P)_{tu}^{rs} \approx 
4\pi\alpha(Q^2)\, D_{\mu\nu}^{\rm free}(p-q)}\\
&& \nonumber
\left[\rule{0mm}{0.7\baselineskip} \sfrac{1}{2}\lambda^a\gamma_\mu\, S^{\rm
free}(q_+) \right]_{tr}
\, \left[ \rule{0mm}{0.7\baselineskip}S^{\rm
free}(q_-)\,\sfrac{1}{2}\lambda^a\gamma_\nu \right]_{su},
\end{eqnarray}
because Eqs.~(\ref{z2mu}), (\ref{z3mu}) and (\ref{z1mu}) yield
\begin{equation}
\label{sumanom}
\frac{2\, \gamma_F}{\beta_1} + \frac{\gamma_1}{\beta_1}  -  \frac{2\,
\gamma_\Gamma}{\beta_1} = 1\,. 
\end{equation}

This is one way of understanding the origin of an often used {\it Ansatz} in
studies of the gap equation; i.e., 
making the replacement
\begin{equation}
\label{abapprox}
g^2 D_{\mu\nu}(k) \to 4\pi\,\alpha(k^2) \,D_{\mu\nu}^{\rm free}(k)
\end{equation}
in Eq.~(\ref{gendse}), and using the ``rainbow truncation:''
\begin{equation}
\label{rainbow}
\Gamma_\nu(q,p)=\gamma_\nu \, .
\end{equation}

Equation~(\ref{abapprox}) is often described as the ``Abe\-lian
approximation'' because the left- and right-hand-sides [r.h.s.] are {\it
equal} in QED.  In QCD, equality between the two sides cannot be obtained
easily by a selective resummation of diagrams.  As reviewed in
Ref.~\cite{cdragw}, Eqs.~(5.1-5.8), it can only be achieved by enforcing
equality between the renormalisation constants for the\linebreak ghost-gluon
vertex and ghost wave function: $\tilde Z_1$ $=$ $\tilde Z_3$.  A mutually
consistent constraint, which follows formally from $\tilde Z_1=\tilde Z_3$,
is to enforce the Abelian Ward identity: $Z_1 = Z_2$.  At one-loop this
corresponds to neglecting the contribution of the 3-gluon vertex to
$\Gamma_\nu$, in which case $\gamma_\Gamma \to \sfrac{2}{3}\xi = \gamma_F$.
This additional constraint provides the basis for extensions of
Eq.~(\ref{rainbow}); i.e., using {\it Ans\"atze} for $\Gamma_\nu$ that are
consistent with the QED vector Ward-Takahashi identity; e.g., Refs.\
\cite{fredthesis}.

Arguments such as these inspire the following {\it Ansatz} for the kernel in
Eq.~(\ref{gendse})~\cite{mr97}:
\begin{eqnarray}
\label{ouransatz}
\lefteqn{Z_1\, \int^\Lambda_q \, g^2 D_{\mu\nu}(p-q)
\frac{\lambda^a}{2}\gamma_\mu 
S(q) \Gamma^a_\nu(q,p)}\\
&&\nonumber \!\!\to \int^\Lambda_q \, {\cal G}((p-q)^2)\, D_{\mu\nu}^{\rm
free}(p-q) \frac{\lambda^a}{2}\gamma_\mu S(q) \frac{\lambda^a}{2}\gamma_\nu
\,,
\end{eqnarray}
with the ultraviolet behaviour of the so-called ``effective coupling:''
${\cal G}(k^2)$, fixed by that of the running strong-coupling.  Since it is
not possible to calculate $Z_1$ nonperturbatively without an\-a\-ly\-sing the
DSE for the dressed-quark-gluon vertex, this {\it Ansatz} absorbs it in the
model effective coupling.

Equation~(\ref{ouransatz}) is a model for the product of the
dressed-propagator and dressed-vertex and its definition is complete once the
behaviour of ${\cal G}(k^2)$ in the infrared is specified; i.e., for $k^2
\lsim 1$-$2\,$GeV$^2$.  Reference~\cite{mr97} used
\begin{eqnarray}
\nonumber\lefteqn{\frac{{\cal G}(k^2)}{k^2} = 8\pi^4 D \delta^4(k) +
\frac{4\pi^2}{\omega^6} D k^2 {\rm e}^{-k^2/\omega^2}}\\
&& \!\!\!\!+ \,\frac{ 8\,\gamma_m\, \pi^2} { \ln\left[\tau + \left(1 +
k^2/\Lambda_{\rm QCD}^2\right)^2\right]} {\cal F}(k^2) \,,
\label{gk2}
\end{eqnarray}
with ${\cal F}(k^2)= [1 - \exp(-k^2/[4 m_t^2])]/k^2$ and $\tau={\rm e}^2-1$.
For $N_f=4$, $\Lambda_{\rm QCD}^{N_f=4}= 0.234\,{\rm GeV}$.  

The qualitative features of Eq.\ (\ref{gk2}) are\linebreak plain.  The first
term is an integrable infrared singularity~\cite{mn83} and the second is a
finite-width approximation to $\delta^4(k)$, normalised such that it has the
same $\int d^4k$ as the first term.  In this way the infrared strength is
split into the sum of a zero-width and a finite-width piece.  The last term
in Eq.~(\ref{gk2}) is proportional to $\alpha(k^2)/k^2$ at large
spacelike-$k^2$ and has no singularity on the real-$k^2$ axis.

There are ostensibly three parameters in\linebreak Eq.~(\ref{gk2}): $D$,
$\omega$ and $m_t$.  However, in Ref.~\cite{mr97}
$\omega=0.3\,$GeV$\,(=1/[.66\,{\rm fm}])$ and $m_t=0.5\,$GeV$\,(=1/[.39\,{\rm
fm}])$ were fixed, and only $D$ and the renormalised $u=d$- and
$s$-current-quark masses were varied in an attempt to obtain a good
description of low-energy $\pi$- and $K$-meson properties, using a
renormalisation point $\zeta=19\,$GeV that is large enough to be in the
perturbative domain.  [The numerical values of $\omega$ and $m_t$ are chosen
so as to ensure that ${\cal G}(k^2)\approx 4\pi \alpha(k^2)$ for $k^2>
2\,$GeV$^2$.  Minor variations in $\omega$ and $m_t$ can be compensated by
small changes in $D$.]  Such a procedure could self-consistently yield $D=0$,
which would indicate that agreement with observable phenomena precludes an
infrared enhancement in the effective interaction.  However, that was not the
case and a good fit required
\begin{equation}
\label{Dvalue}
D= (0.884\,{\rm GeV})^2\,,
\end{equation}
with renormalised current-quark masses
\begin{equation}
\label{params}
\begin{array}{cc}
m_{u,d}(\zeta) = 3.74\,{\rm MeV}\,,\; &
m_s(\zeta) = 82.5\,{\rm MeV}\,,
\end{array}
\end{equation}
which are in the ratio $1\,$:$\,22$, and yielded, in MeV,
\begin{equation}
\begin{array}{l|cccc}
         & m_\pi & m_K & f_\pi & f_K \\\hline
{\rm Calc.\,~\protect\cite{mr97}} & 139 & 497 & 131 & 154 \\
{\rm Expt.~\protect\cite{pdg98}} & 139 & 496 & 131 & 160
\end{array}
\end{equation}
and other quantities to be described below.  An explanation of how this fit
was accomplished requires a discussion of the homogeneous Bethe-Salpeter
equation, which I postpone until Sec.\ \ref{sec.bound}.  It is described in
detail in Refs.\ \cite{mr97,pieterVM}.  Here I focus instead on describing
the properties of the DSE solution obtained with these parameter values.
\label{bsefit}

Using Eqs.~(\ref{gendse}-\ref{mzeta}) and (\ref{ouransatz}) the gap equation
can be written
\begin{equation}
\label{dsemod}
S(p,\zeta)^{-1} = 
Z_2\, i\gamma\cdot p + Z_4\, m(\zeta) + \Sigma^\prime
(p,\Lambda)\,,
\end{equation}
with the regularised quark self energy
\begin{eqnarray}
\label{sigmod}
\lefteqn{\Sigma^\prime(p,\Lambda)  := }\\
&& \nonumber \int^\Lambda_q \, {\cal G}((p-q)^2)\,
D_{\mu\nu}^{\rm free}(p-q) \frac{\lambda^a}{2}\gamma_\mu S(q)
\frac{\lambda^a}{2}\gamma_\nu \,.
\end{eqnarray}
When $\hat m \neq 0$ the renormalisation condition,\linebreak
Eq.~(\ref{renormS}), is straightforward to implement.  Writing
\begin{equation}
\Sigma^\prime(p,\Lambda) := i \gamma\cdot p \, \left(
                A^\prime(p^2,\Lambda^2) - 1\right) +
                B^\prime(p^2,\Lambda^2)\,,
\end{equation}
which emphasises that these functions depend on the regularisation
mass-scale, $\Lambda$, Eq.~(\ref{renormS}) entails
\begin{equation}
\label{z2def}
Z_2(\zeta^2,\Lambda^2)  =  2 - A^\prime(\zeta^2,\Lambda^2)\,,
\end{equation}
\begin{equation}
m(\zeta) =  Z_2(\zeta^2,\Lambda^2)\,m_{\rm bm}(\Lambda^2) + 
        B^\prime(\zeta^2,\Lambda^2)
\end{equation}
so that
\begin{equation}
\label{arenbren}
A(p^2,\zeta^2)   =  1 
        +  A^\prime(p^2,\Lambda^2) 
        - A^\prime(\zeta^2,\Lambda^2)\,,
\end{equation}
\begin{equation}
B(p^2,\zeta^2)   =  m(\zeta) 
        +  B^\prime(p^2,\Lambda^2) 
        - B^\prime(\zeta^2,\Lambda^2)\,.
\end{equation}

Multiplicative renormalisability requires that having fixed the solutions at
a single renormalisation point, $\zeta$, their form at another point,
$\tilde\zeta$, is given by
\begin{eqnarray}
S^{-1}(p,\tilde\zeta) & = &
        \frac{Z_2(\tilde\zeta^2,\Lambda^2)}{Z_2(\zeta^2,\Lambda^2)}
        S^{-1}(p,\zeta)\,.
\end{eqnarray}
This feature is evident in the solutions obtained in Ref.~\cite{mr97}.  It
means that, in evolving the renormalisation point to $\tilde\zeta$, the ``1''
in Eqs.~(\ref{arenbren}) is replaced by
$Z_2(\tilde\zeta^2,\Lambda^2)/Z_2(\zeta^2,\Lambda^2)$, and the ``$m(\zeta)$''
by $m(\tilde\zeta)$; i.e., the ``seeds'' in the integral equation evolve
according to the QCD renormalisation group. This is why Eq.~(\ref{ouransatz})
is called a ``renormalisation-group-improved rainbow truncation.''

Returning to the chiral limit, it follows from Eqs.~(\ref{Zmass}),
(\ref{mzeta}), (\ref{masanom}) and (\ref{chirallimit}) that for $\hat m = 0$
\begin{equation}
\label{chiralA}
Z_2(\zeta^2,\Lambda^2) \,m_{\rm bare}(\Lambda^2)=0\,,\;\forall \Lambda\,.
\end{equation}
Hence, as remarked on page~\pageref{`without subtraction'}, there is no
subtraction in the equation for $B(p^2,\zeta^2)$; \linebreak i.e.,
Eq.~(\ref{arenbren}) becomes
\begin{equation}
B(p^2,\zeta^2)  =  B^\prime(p^2,\Lambda^2) \,,\;
\lim_{\Lambda\to \infty} B^\prime(p^2,\Lambda^2) < \infty\,,
\end{equation}
which is only possible if the mass function is at least $1/p^2$-suppressed.
This is not the case in quenched strong-coupling QED, where the mass function
behaves as~\cite{fukudacahillQED}
\begin{equation}
\propto \cos({\rm const.}\,\ln
[p^2/\zeta^2])/(p^2/\zeta^2)^{1/2}\,,
\end{equation}
and that is the origin of the complications discussed in
Refs.~\cite{cdragw,sizer,manuel}.

\FIGURE[hbt]{\epsfig{figure=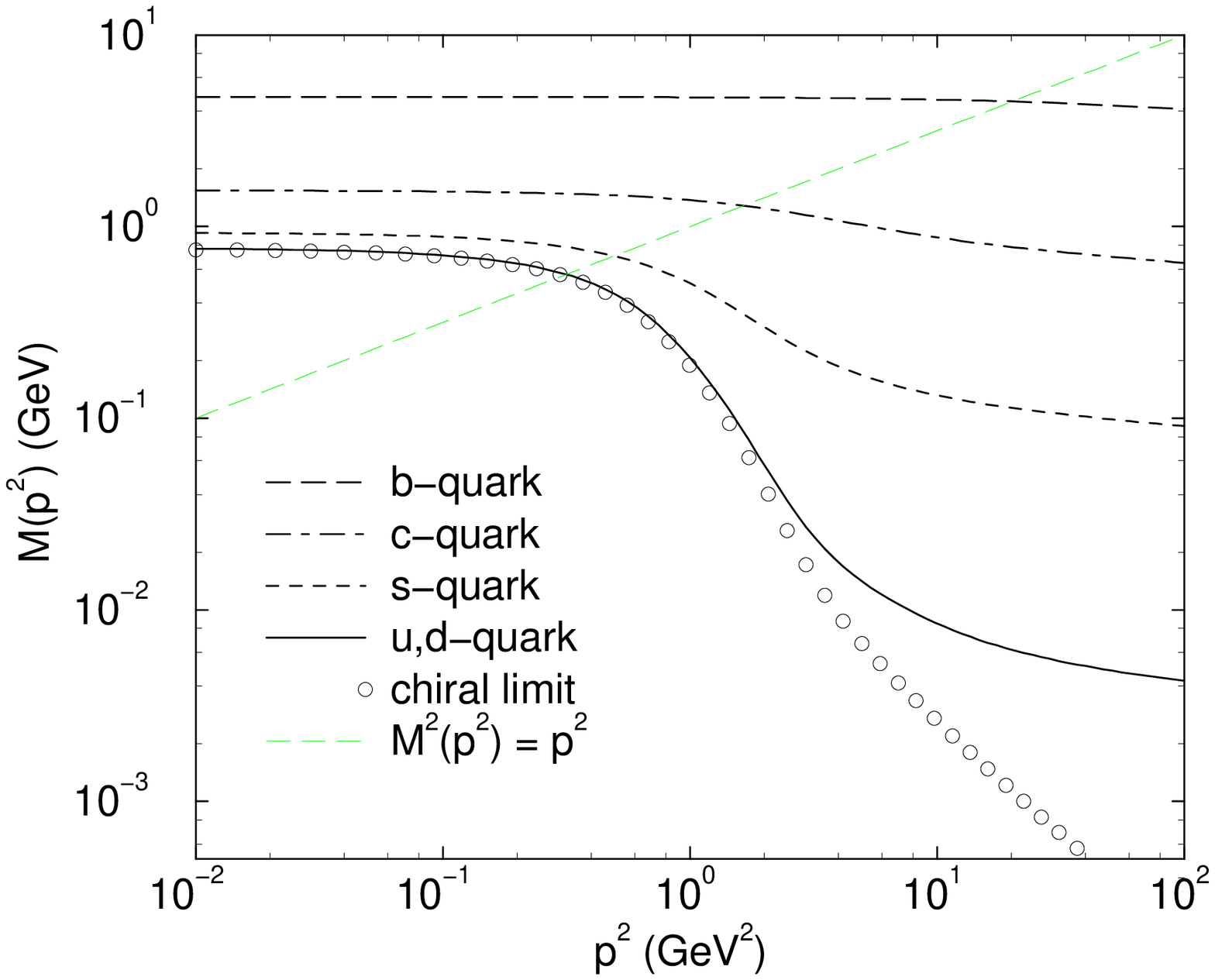,height=5.9cm}\caption{Quark mass function
obtained as a solution of Eq.~(\protect\ref{gendse}) using the model of
Eqs.~(\protect\ref{ouransatz}), (\protect\ref{gk2}), and current-quark
masses, fixed at a renormalisation point $\zeta= 19\,$GeV: $m_{u,d}^\zeta =
3.7\,$MeV, $m_s^\zeta = 82\,$MeV, $m_c^\zeta=0.58\,$GeV and
$m_b^\zeta=3.8\,$GeV.  The indicated solutions of $M^2(p^2)=p^2$ define the
Euclidean constituent-quark mass, $M^E_f$, which takes the values:
$M^E_u=0.56\,$GeV, $M^E_s=0.70\,$GeV, $M^E_c= 1.3\,$GeV, $M^E_b= 4.6\,$GeV.
These values and their ratios are consistent with contemporary phenomenology;
e.g., Refs.\protect\cite{pdg98,simon}\label{mp2fig} (Figure adapted from
Refs.~\protect\cite{pieterrostock,mishaSVY}.)
\label{massfunction}}}

In Fig.~\ref{massfunction} I present the renormalised dressed-quark mass
function, $M(p^2)$, obtained by solving Eq.~(\ref{dsemod}) using the model
and parameter values of Ref.~\cite{mr97},
Eqs.~(\ref{ouransatz}-\ref{params}), and also in the chiral limit and with
typical heavy-quark current-mass values.

In the presence of explicit chiral
symmetry breaking Eq.~(\ref{masanom}) describes the form of $M(p^2)$ for $p^2
> {\rm O}(1\,{\rm GeV}^2)$.  In the chiral limit, however, the ultraviolet
behaviour is given by
\begin{eqnarray}
\label{Mchiral}
\lefteqn{M(p^2) \stackrel{{\rm large}-p^2}{=}\,}\\
&& \nonumber 
\frac{2\pi^2\gamma_m}{3}\,\frac{\left(-\,\langle \bar q q \rangle^0\right)}
           {p^2
        \left(\sfrac{1}{2}\ln\left[p^2/\Lambda_{\rm QCD}^2\right]
        \right)^{1-\gamma_m}}\,,
\end{eqnarray}
where $\langle \bar q q \rangle^0$ is the
renormalisation-point-in\-de\-pen\-dent vacuum quark condensate.  This
behaviour too is characteristic of the QCD renormalisation group~\cite{hdp76}
and exhibits the power-law suppression anticipated on page~\pageref{`without
subtraction'}.  These results for the large-$p^2$ behaviour of the mass
function are model independent; i.e., they arise only because the DSE
truncation is consistent with the QCD renormalisation group at one loop.  (It
has long been known that the truncation defined by\linebreak
Eq.~(\ref{ouransatz}) yields results in agreement with the QCD
renormalisation group at one loop; e.g.,
Refs.~\cite{higashijimaportermelbourne}.)

The gauge invariant expression for the
re\-nor\-ma\-li\-sa\-tion-point-dependent vacuum quark condensate was derived
in Ref.~\cite{mrt98}:
\begin{equation}
\label{qbq0}
\,-\,\langle \bar q q \rangle_\zeta^0 := 
Z_4(\zeta^2,\Lambda^2)\, N_c {\rm tr}_{\rm D}\int^\Lambda_q\,
        S^{0}(q,\zeta)\,,
\end{equation}
where ${\rm tr}_D$ identifies a trace over Dirac indices only and the
superscript ``$0$'' indicates the quantity was calculated in the chiral
limit.  Substituting Eq.~(\ref{Mchiral}) into Eq.~(\ref{qbq0}), recalling
that $Z_4=Z_m$ in Landau gauge and using Eq.~(\ref{zmdef}) leads to the
one-loop expression
\begin{equation}
\label{qbqzeta}
\langle \bar q q \rangle_\zeta^0 = \left(\sfrac{1}{2}\ln \zeta^2/\Lambda_{\rm
QCD}^2\right)^{\gamma_m} \, \langle \bar q q \rangle^0\,.
\end{equation}
Employing Eq.~(\ref{masanom}), this exemplifies the general result that
\begin{equation}
\label{mqbq}
m(\zeta)\,\langle \bar q q \rangle_\zeta^0 = 
\hat m \langle \bar q q \rangle^0\,;
\end{equation}
i.e., that this product is renormalisation point invariant and, importantly,
it shows that the behaviour expressed in Eq.~(\ref{Mchiral}) is exactly that
required for consistency with the gauge invariant expression for the quark
condensate.  A model, such as Ref.~\cite{diakonov}, in which the scalar
projection of the chiral limit dressed-quark propagator falls faster than
$1/p^4$, up to $\ln$-corrections, is only consistent with this quark
condensate vanishing, and it is this condensate that appears in the current
algebra expression for the pion mass~\cite{mrt98}, as discussed in connection
with Eq.\ (\ref{gmor}).

Equation~(\ref{Mchiral}) provides a reliable means of calculating the quark
condensate because corrections are suppressed by powers of $\Lambda_{\rm
QCD}^2/\zeta^2$.  Analysing the asymptotic form of the numerical solution one
finds
\begin{equation}
\label{qbqM0}
-\langle \bar q q \rangle^0 = (0.227\,{\rm GeV})^3\,.
\end{equation}
Using Eq.~(\ref{qbqzeta}) one can define a one-loop evolved condensate
\begin{equation}
\label{qbq1}
\left.-\langle \bar q q \rangle_\zeta^0\right|_{\zeta=1\,{\rm GeV}}  := 
-\left(\ln\left[1/\Lambda_{\rm QCD}\right]\right)^{\gamma_m} \, \langle \bar
q q \rangle^0\,,
\end{equation}
which takes the value 
\begin{equation}
\left.-\langle \bar q q \rangle_\zeta^0\right|_{\zeta=1\,{\rm GeV}} =
(0.241\,{\rm GeV})^3\,.
\end{equation}
This can be directly compared with the value of the quark condensate employed
in contemporary phenomenological studies~\cite{derek}: 
\begin{equation}
(0.236\pm 0.008\,{\rm GeV})^3\,.
\end{equation}
It was noted in Ref.~\cite{mr97} that increasing $\omega$ $\to$ $1.5\,\omega$
in ${\cal G}(k^2)$ increases the calculated value in Eq.~(\ref{qbq1}) by
$\sim 10$\%; i.e., the magnitude of the condensate is correlated with the
degree of infrared enhancement/strength in the effective interaction.  That
is unsurprising because it has long been known that there is a critical
coupling for DCSB; i.e., the kernel in the gap equation must have an
integrated strength that exceeds some critical
value~\cite{higashijimaportermelbourne}.  This is true in all fermion-based
studies of DCSB, a point discussed in more detail on page
\pageref{sigmadelta}.

The renormalisation-point-invariant current-quark masses corresponding to the
$m_f(\zeta)$ in\linebreak Fig.~\ref{massfunction} are obtained in the
following way: using Eq.~(\ref{qbq0}), direct calculation from the chiral
limit numerical solution gives
\begin{equation}
\langle \bar q q \rangle_{\zeta=19\,{\rm GeV}}^0 = - (0.275\,{\rm GeV})^3\,,
\end{equation}
and hence from the values of $m_f^\zeta \equiv m_f(\zeta)$ listed in
Fig.~\ref{massfunction} and Eqs.~(\ref{mqbq}), (\ref{qbqM0}), in MeV,
\begin{equation}
\begin{array}{cc}
\hat m_{u,d}=6.60\,,\, &
\hat m_s = 147\,,\, \\
\hat m_c = 1\,030\,,\, &
\hat m_b = 6\,760\,,
\end{array} 
\end{equation}
from which also follow one-loop evolved values in analogy with
Eq.~(\ref{qbq1}):
\begin{equation}
\label{oneloopmasses}
\begin{array}{cc}
m_{u,d}^{1\,{\rm GeV}}= 5.5\,,\, &
m_s^{1\,{\rm GeV}} = 130\,,\, \\
m_c^{1\,{\rm GeV}} = 860\,,\, &
m_b^{1\,{\rm GeV}} = 5\,700\,.
\end{array} 
\end{equation}
Smaller values of the light-current-quark masses would require a larger value
of the vacuum quark condensate in order to be consistent with light-meson
properties.

Figure~\ref{massfunction} highlights a number of qualitative aspects of the
quark mass function.  One is the difference in the ultraviolet between the
be\-ha\-vi\-our of $M(p^2)$ in the chiral limit and in the presence of
explicit chiral symmetry breaking.  In the infrared, however, the $u,d$-quark
mass function and the chiral limit solution are almost indistinguishable.
The chiral limit solution is nonzero {\it only} because of the
nonperturbative DCSB mechanism whereas the $u,d$-quark mass function is
purely perturbative at $p^2>20\,$GeV$^2$.  Hence the evolution to coincidence
of the chiral-limit and $u,d$-quark mass functions makes clear the transition
from the perturbative to the nonperturbative domain.

\begin{figure}[ht]
\centering{\ \epsfig{figure=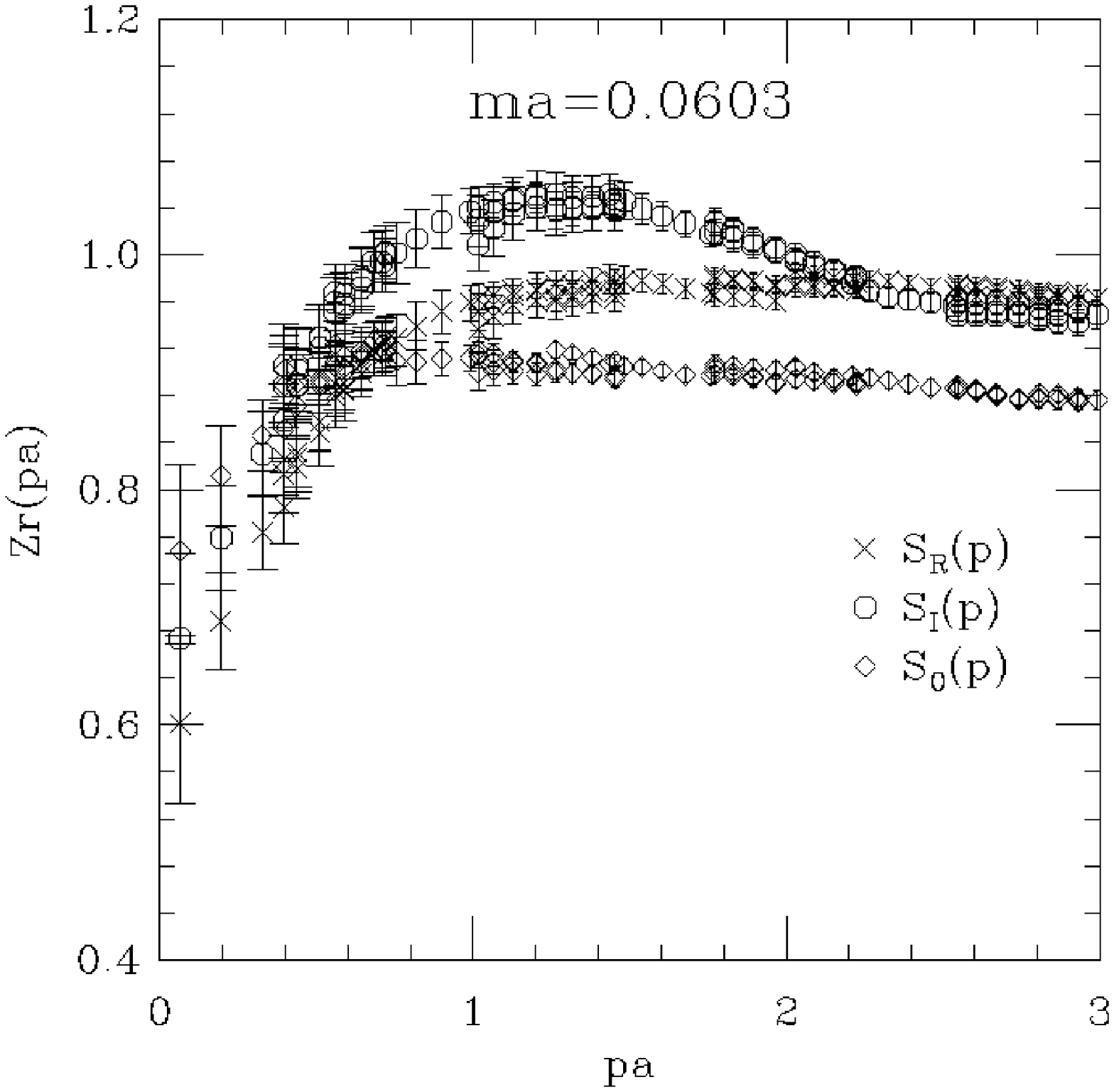,height=6.5cm} }
\vspace*{-6.1cm}

\hspace*{0.54cm}\epsfig{figure=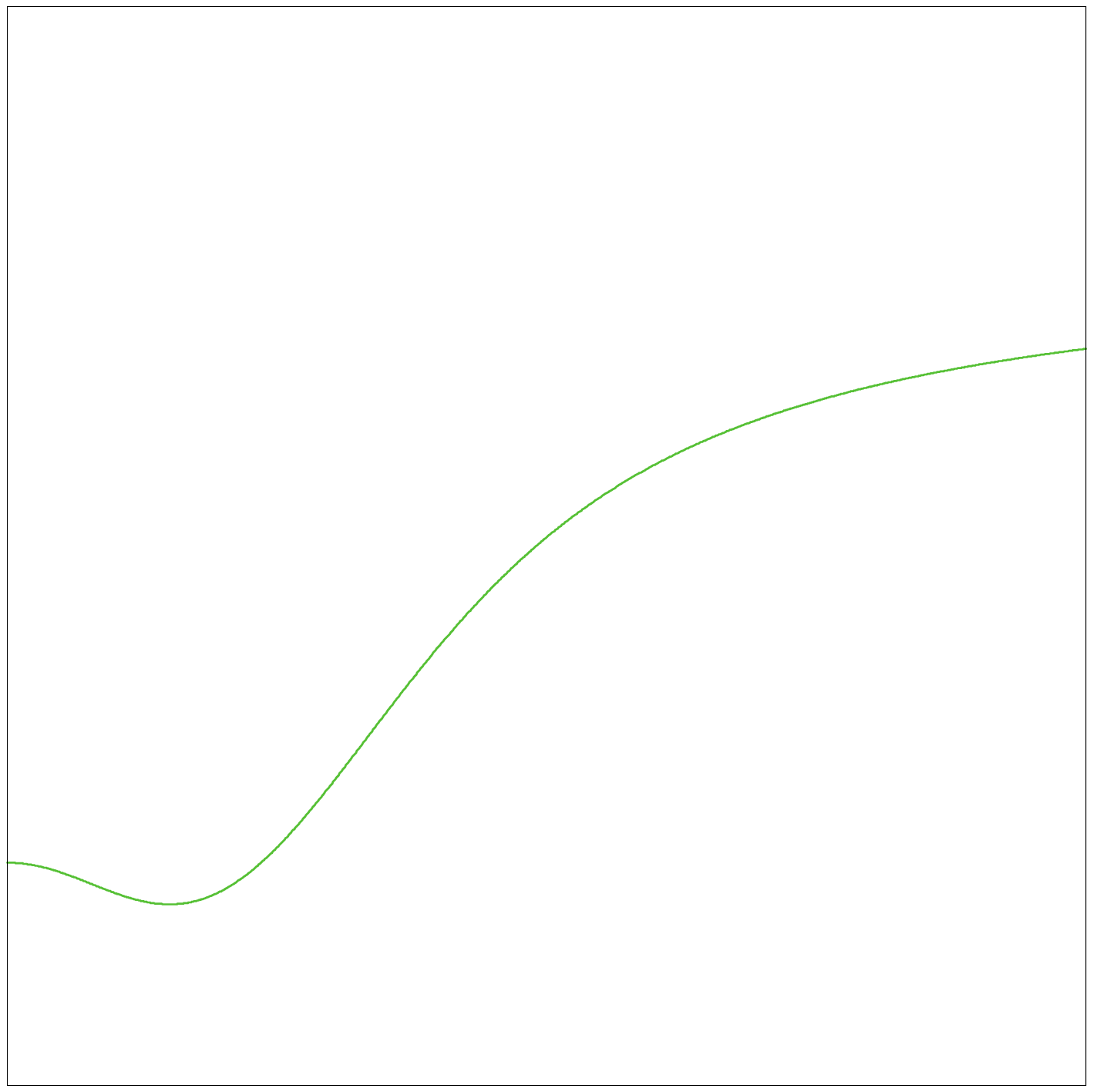,height=5.3cm}\vspace*{0.7cm}

\centering{\ \epsfig{figure=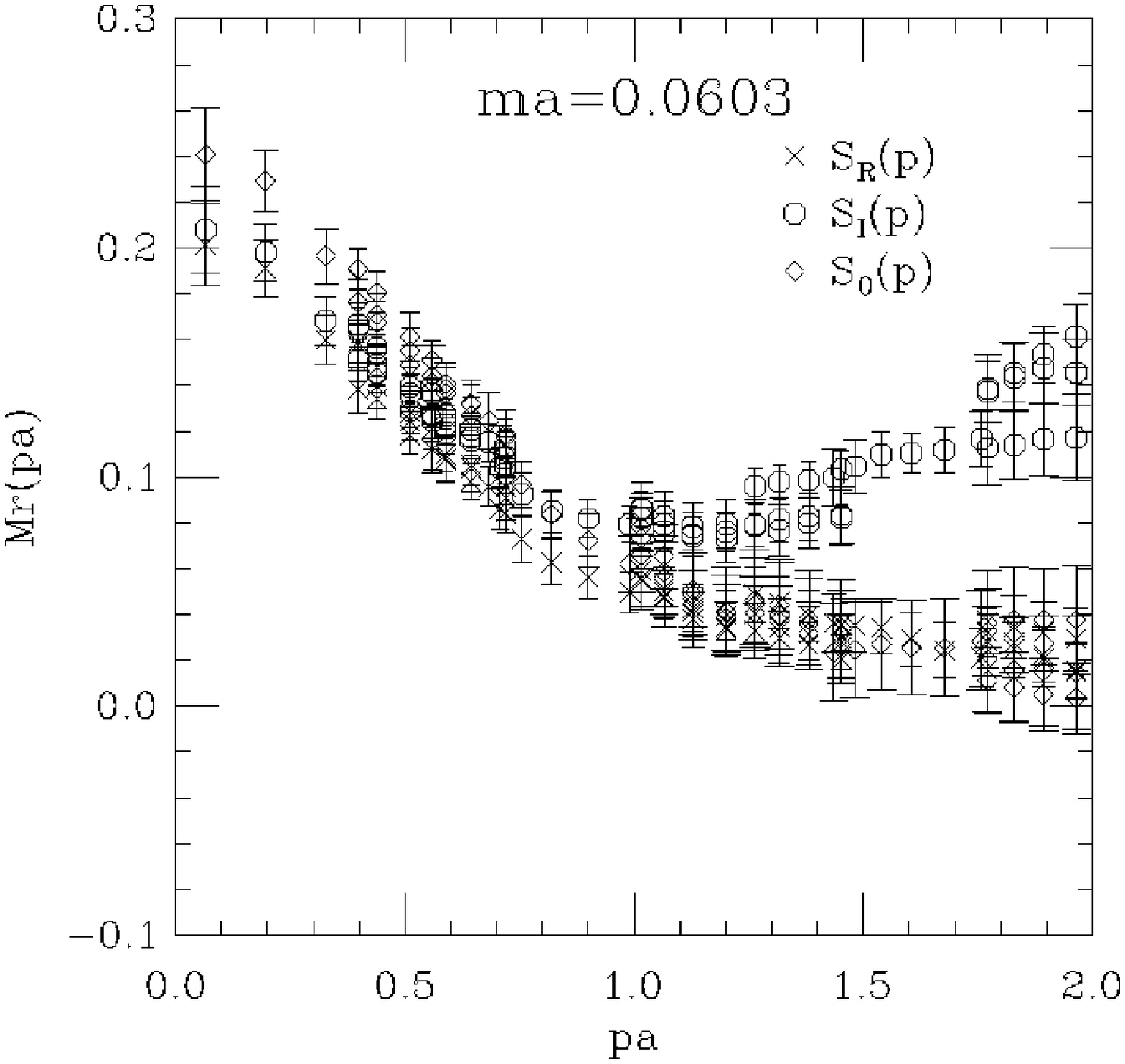,height=6.5cm} }
\vspace*{-6.2cm}

\hspace*{0.40cm}\epsfig{figure=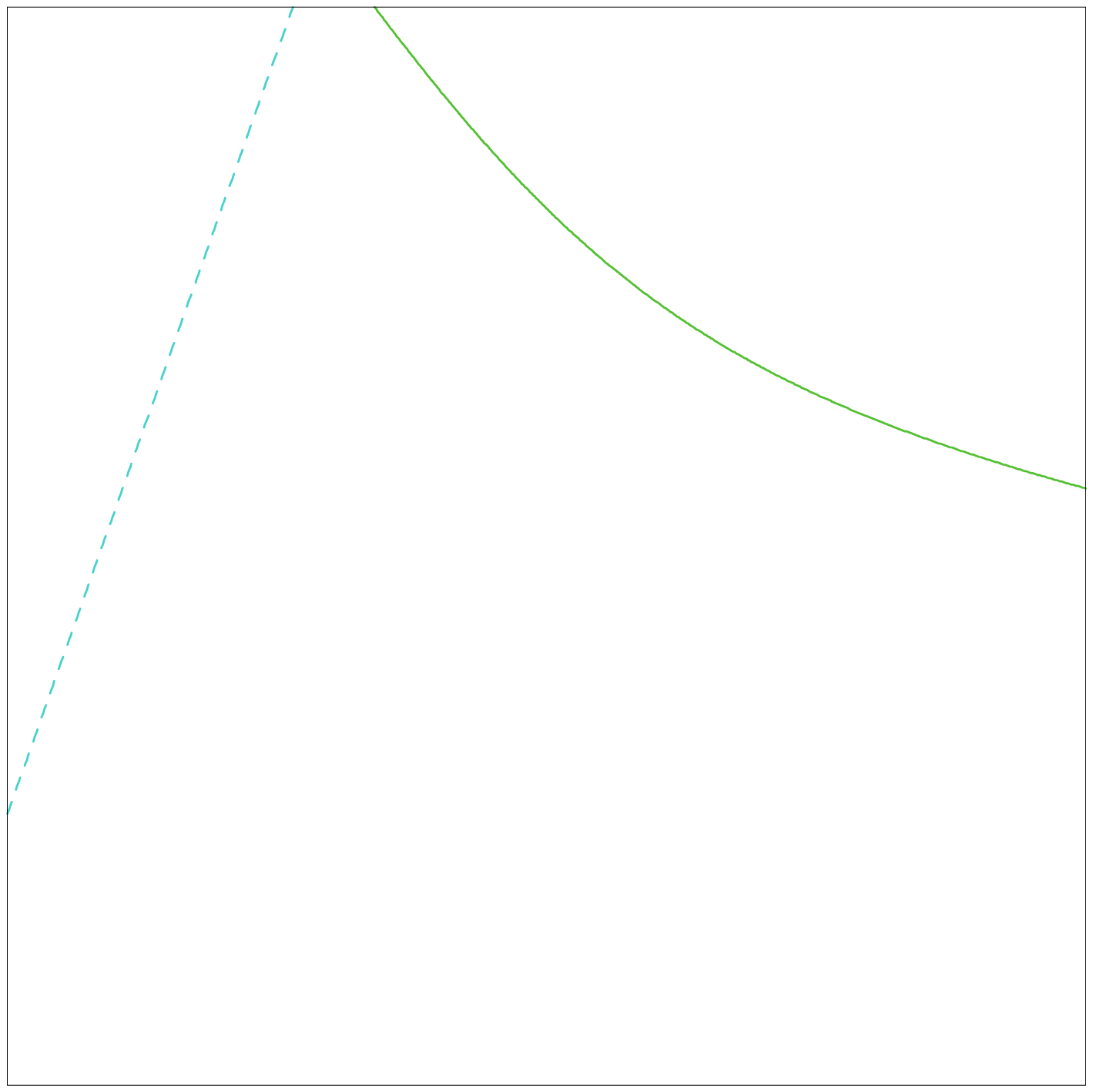,height=5.3cm}\vspace*{0.5cm}

\caption{\label{latZZ} {\it Upper Panel}: Lattice result for $Z(pa)$,
$a\simeq 2.0\,\mbox{GeV}^{-1}$ is the lattice spacing, calculated with
$ma=0.0603$~\protect\cite{tonylatticequark}, compared with the DSE analogue:
$Z(x)$, $x$ is a dimensionless momentum variable, from the phenomenological
algebraic model of Ref.~\protect\cite{myriad}.  The variation between the
lattice results is an indication of extant numerical artefacts.  The ``dip''
in $Z(x)$ for $x\simeq 0$ is a long-standing prediction of phenomenologically
efficacious DSE studies.
{\it Lower Panel}: Analogous plot for the mass function.  Numerical artefacts
are more significant in this case.  Nevertheless the enhancement at
small-$pa$, indicative of DCSB, is clearly evident and in semi-quantitative
agreement with that predicted by DSE studies.  The dashed-line assists with
estimating the Euclidean constituent-quark mass: $\sim 200\,$MeV in the
lattice simulation and $\sim 350$ in the phenomenological DSE
model~\protect\cite{myriad}.}
\end{figure}

It is on this nonperturbative domain that $A(p^2,\zeta^2)$ differs
significantly from one.  A concomitant observation is that the DCSB mechanism
has a significant effect on the propagation characteristics of
$u,d,s$-quarks.  These aspects of the momentum-dependence of the scalar
functions in the dressed-light-quark propagators have recently been confirmed
in numerical simulations of lattice-QCD~\cite{tonylatticequark}, as
illustrated in the comparisons of Fig.~\ref{latZZ}.

Returning to Fig.~\ref{massfunction}, it is evident that the large
current-quark mass of the $b$-quark almost entirely suppresses
momentum-de\-pen\-dent dressing, so that $M_b(p^2)$ is nearly constant on a
substantial domain.  This is true to a lesser extent for the $c$-quark but in
both cases one observes that DCSB is not the dominant mass-generating
mechanism.  

A means of quantifying the effect of the\linebreak DCSB mechanism on
dressed-quark propagation characteristics is now obvious.  One can introduce
\cite{pieterrostock,mishaSVY}: ${\cal L}_f:=M^E_f/m_f^\zeta$, where $M^E_f$
is the Euclidean con\-sti\-tu\-ent-quark mass, defined in
Fig.~\ref{massfunction} as the solution of
\begin{equation}
\label{euclidmass}
(M_f^E)^2 - M^2(p^2=(M_f^E)^2,\zeta^2) = 0\,,
\end{equation}
and this ratio takes the values
\begin{equation}
\label{Mmratio}
\begin{array}{c|ccccc}
        f   &   u,d  &   s   &  c  &  b  \\ \hline
 {\cal L}_f &  150   &    10  &  2.2 &  1.2 
\end{array}\,.
\end{equation}

The values are representative and definitive: for light-quarks ${\cal
L}_{q=u,d,s} \sim 10$-$100$, while for\linebreak heavy-quarks ${\cal
L}_{Q=c,b} \sim 1$; and highlight the existence of a mass-scale
characteristic of DCSB: $M_\chi$.  The propagation characteristics of a
flavour with\linebreak $m_f^\zeta\leq M_\chi$ are significantly altered by
the DCSB mechanism, while for flavours with $m_f^\zeta\gg M_\chi$
mo\-men\-tum-dependent dressing is almost irrelevant.  It is apparent and
unsurprising that $M_\chi \sim 0.2\,$GeV$\,\sim \Lambda_{\rm QCD}$.  This
feature of the dressed-quark mass function provides the foundation for a
constituent-quark-like approximation in the\linebreak treatment of
heavy-meson decays and transition form factors~\cite{mishaSVY,usPLB}.
 
\section{Confinement}
\label{sec.conf}
To proceed it is necessary to discuss confinement and I begin with a
definition.  Confinement is the failure to directly observe coloured
excitations in a detector: neither quarks nor gluons nor coloured composites.
The contemporary hypothesis is stronger; i.e., coloured excitations cannot
propagate to a detector.  To ensure this it is sufficient that coloured
$n$-point functions violate the axiom of reflection positivity~\cite{gj81},
which is guaranteed if the Fourier transform of the momentum-space $n$-point
Schwinger function is not a positive-definite function of its arguments.

Reflection positivity is one of a set of five axioms that must be satisfied
if the given $n$-point function is to have a continuation to Min\-kow\-ski
space and hence an association with a physical, observable state.  If an
Hamiltonian exists for the theory but a given $n$-point function violates
reflection positivity then the space of observable states, which is spanned
by the eigenstates of the Hamiltonian, does not contain anything
corresponding to the excitation(s) described by that Schwinger function.

The free boson propagator does not violate reflection positivity:
\begin{eqnarray}
\label{freeboson}
\Delta(x) & := &
\int\frac{d^4 k}{(2\pi)^4}\,{\rm e}^{i k\cdot x}\,\frac{1}{k^2+m^2}\\
& = & \frac{m}{4\pi^2 x}\,K_1(m x)\,.
\end{eqnarray}
Here $x:= (x\cdot x)^{1/2}>0$ and $K_1$ is the monotonically decreasing,
strictly convex-up, non-negative modified Bessel function of the second kind.
The same is true of the free fermion propagator:
\begin{eqnarray}
\label{freefermion}
\lefteqn{S(x) = \int\frac{d^4 k}{(2\pi)^4}\,{\rm e}^{i k\cdot x}\,
\frac{m-i\gamma\cdot k }{k^2+m^2}}\\
&  & =\frac{m^2}{4\pi^2 x} \left[ K_1(m x) + \frac{\gamma\cdot x }{x} K_2(m
x)\right]\,,
\end{eqnarray}
which is also positive definite.  

The spatially averaged Schwinger function is a particularly insightful
tool~\cite{fredIR,hollenberg}.  Consider the fermion and let $T=x_4$
represent Euclidean ``time,'' then
\begin{eqnarray}
\label{sigmasT}
\sigma_S(T)&:=& \int d^3 x\, \sfrac{1}{4}{\rm tr}_D S(\vec{x},T) \nonumber \\
& = &\frac{1}{\pi}\int_0^{\infty}\,d\ell\,\frac{m}{\ell^2+m^2}\,\cos(\ell T)
\nonumber \\ &= & \sfrac{1}{2}\,{\rm e}^{-m T}.
\end{eqnarray}
Hence the free fermion's mass can easily be obtained from the large $T$
behaviour of the spatial average:
\begin{equation}
\label{mTfree}
m \,T = - \lim_{T\to \infty} \ln \sigma_S(T)\,.
\end{equation}
[The boson analogy is obvious.]  This is just the approach used to determine
bound state masses in simulations of lattice-QCD.

For contrast, consider the dressed-gluon $2$-point function obtained
with~\cite{stingl123}
\begin{equation}
\label{vanishing}
d(k^2) = \frac{k^4}{k^4 + \gamma^4}
\end{equation}
in Eq.~(\ref{gluoncovariant})
\begin{eqnarray}
\label{badDx}
D(x)& := & \int\frac{d^4 k}{(2\pi)^4}\,{\rm e}^{i k\cdot x}\, \frac{k^2}{k^4 +
\gamma^4} \\
& = & -\frac{\gamma}{4\pi^2 x} \, 
\left.\left(\frac{d}{dz}{\rm  ker}(z)\right)\right|_{z= \gamma x}\!,
\end{eqnarray}
where ker$(z)$ is the oscillatory Thomson function.  $D(x)$ is not positive
definite and hence a dressed-gluon $2$-point function that vanishes at
$k^2=0$ violates the axiom of reflection positivity and is therefore not
observable; i.e., the excitation it describes is confined.  

At asymptotically large Euclidean distances
\begin{eqnarray}
\lefteqn{D(x)  \stackrel{x\to\infty}{\propto}  \frac{\gamma^{1/2}}{x^{3/2}}\,
{\rm e}^{-\gamma x/\surd 2}}\\
& & \nonumber \times
\left[\cos(\sfrac{1}{\sqrt{2}}\gamma x + \sfrac{\pi}{8}) 
+ \sin(\sfrac{1}{\sqrt{2}}\gamma x+ \sfrac{\pi}{8}) \right]\,.
\end{eqnarray}
Comparing this with Eq.~(\ref{freeboson}) one identifies a mass as the
coefficient in the exponential: 
\begin{equation}
m_D =\gamma/\surd 2.
\end{equation}
[NB.\ At large $x$, $K_1(x) \propto \exp(-x)/\surd x$.]  By an obvious
analogy, the coefficient in the oscillatory term is the {\it
lifetime}~\cite{stingl123}:
\begin{equation}
\tau = 1/m_D.
\end{equation}
Both the mass and lifetime are tied to the dynamically generated mass-scale
$\gamma$, which, using
\begin{equation}
\frac{z}{z^2+\gamma^4} = \sfrac{1}{2}\, \frac{1}{z+i\gamma^2} +
\sfrac{1}{2}\,\frac{1}{z-i\gamma^2}\,,
\end{equation}
is just the displacement of the complex conjugate poles from the real-$k^2$
axis.  

It is a general result that the Fourier transform of a real function with
complex conjugate poles is not positive definite.  Hence the existence of
such poles in a $n$-point Schwinger function is a sufficient condition for
the violation of reflection positivity and thus for confinement.  

The spatially averaged Schwinger function is also useful here.
\begin{eqnarray}
\label{D(T)}
\lefteqn{D(T) := \int d^3 x\, D(\vec{x},T) }\nonumber \\
&& = \frac{1}{2\gamma}\,{\rm e}^{-\frac{1}{\surd 2} \gamma T}\,
\cos(\sfrac{1}{\surd 2}\gamma T + \sfrac{\pi}{4}) \,,
\end{eqnarray}
and, generalising Eq.~(\ref{mTfree}), one can define a $T$-dependent mass:
\begin{eqnarray}
\lefteqn{m(T)\,T := -\ln D(T) } \\
\label{oscexamp}
&& = 
 \ln (2 \gamma) + \sfrac{1}{\surd 2} \gamma \,T 
- \ln\left[\cos(\sfrac{1}{\surd 2}\gamma T + \sfrac{\pi}{4})\right] \,.
\nonumber
\end{eqnarray}
It exhibits periodic singularities whose frequency is proportional to the
dynamical mass-scale that is responsible for the violation of reflection
positivity.  If a dressed-fermion $2$-point function has complex conjugate
poles it too will be characterised by a $T$-dependent mass that exhibits such
behaviour.

\subsection{Three-dimensional QED}
\label{sub.qed3}
This reflection positivity criterion has been employed to very good effect in
three dimensional QED~\cite{pieterQED3}.  First, some background.  QED$_3$ is
confining in the quenched truncation~\cite{mack}.  That is evident in the
classical potential
\begin{eqnarray}
V(r)& :=& \int_{-\infty}^{\infty}dx_3\,\int\frac{d^3 k}{(2\pi)^3}\, {\rm
e}^{i\vec{k}\cdot\vec{x} + ik_3 x_3}\,\frac{e^2}{k^2} \nonumber \\
& = &
\frac{e^2}{2\pi}\ln(e^2 r)\,,\;r^2=x_1^2+x_2^2\,,
\label{Vrquench}
\end{eqnarray}
which describes the interaction between two static sources.  [NB.\ $e^2$ has
the dimensions of mass in QED$_3$.]  It is a logarithmically growing
potential, showing that the energy required to separate two charges is
infinite.  Furthermore, $V(r)$ is just a one-dimensional average of the
spatial gauge-boson $2$-point Schwinger function and it is not positive
definite, which indicates that the photon is also confined.

However, if now the photon vacuum polarisation tensor is evaluated at order
$e^2$ using $N_f$ massless fermions then, with the notation of\linebreak
Eq.~(\ref{gluoncovariant}), the photon propagator is characterised
by~\cite{appel}
\begin{equation}
\label{Vrunquench0}
\frac{d(k^2)}{k^2} = \frac{1}{k^2+ \tilde\alpha k}\,,\;\mbox{from~}
\Pi(k^2) = \frac{\tilde\alpha}{k}\,,
\end{equation}
$\tilde\alpha = N_f e^2 /8$, and one finds~\cite{conrad}
\begin{equation}
V(r) = \frac{e^2}{4}
\left[\mbox{\bf H}_0(\tilde\alpha r) - N_0(\tilde\alpha r)\right]\,,
\end{equation}
where {\bf H}$_0(x)$ is a Struve function and $N_0(x)$ a Neumann function,
both of which are related to Bessel functions.  In this case $V(r)$ is
positive definite, with the limiting cases
\begin{equation}
V(r) \stackrel{r \approx 0}{\sim} - \ln(\tilde\alpha r)\,,\;\;
V(r) \stackrel{r \to \infty}{=} \frac{e^2}{2\pi} \,\frac{1}{\tilde\alpha r}\,,
\end{equation}
and confinement is lost in QED$_3$.  That is easy to understand: pairs of
massless fermions cost no energy to produce and can propagate to infinity so
they are very effective at screening the interaction.

With $d(k^2)=1/[1+\Pi(k^2)]$ and sensible, physical constraints on the form
of $\Pi(k^2)$, such as boundedness and vanishing in the ultraviolet, one can
show that~\cite{justin}
\begin{equation}
\label{Vrqed3}
V(r) \stackrel{r \to \infty}{=} \frac{e^2}{2\pi}\,
\frac{1}{1+\Pi(0)}\ln( e^2 r) + {\rm const.} + h(r)\,,
\end{equation}
where $h(r)$ falls-off at least as quickly as $1/r$.  Hence, the existence of
a confining potential in QED$_3$ just depends on the value of the vacuum
polarisation at the origin.  In the quenched truncation, $\Pi(0) = 0$ and the
theory is logarithmically confining.  With massless fermions,\linebreak
$1/[1+\Pi(0)]=0$ and confinement is absent.  Finally, when the vacuum
polarisation is evaluated from a loop of massive fermions, whether that mass
is obtained dynamically via the gap equation or simply introduced as an
external parameter, one obtains $\Pi(0) < \infty$ and hence a confining
theory.

\FIGURE[hbt]{\epsfig{figure=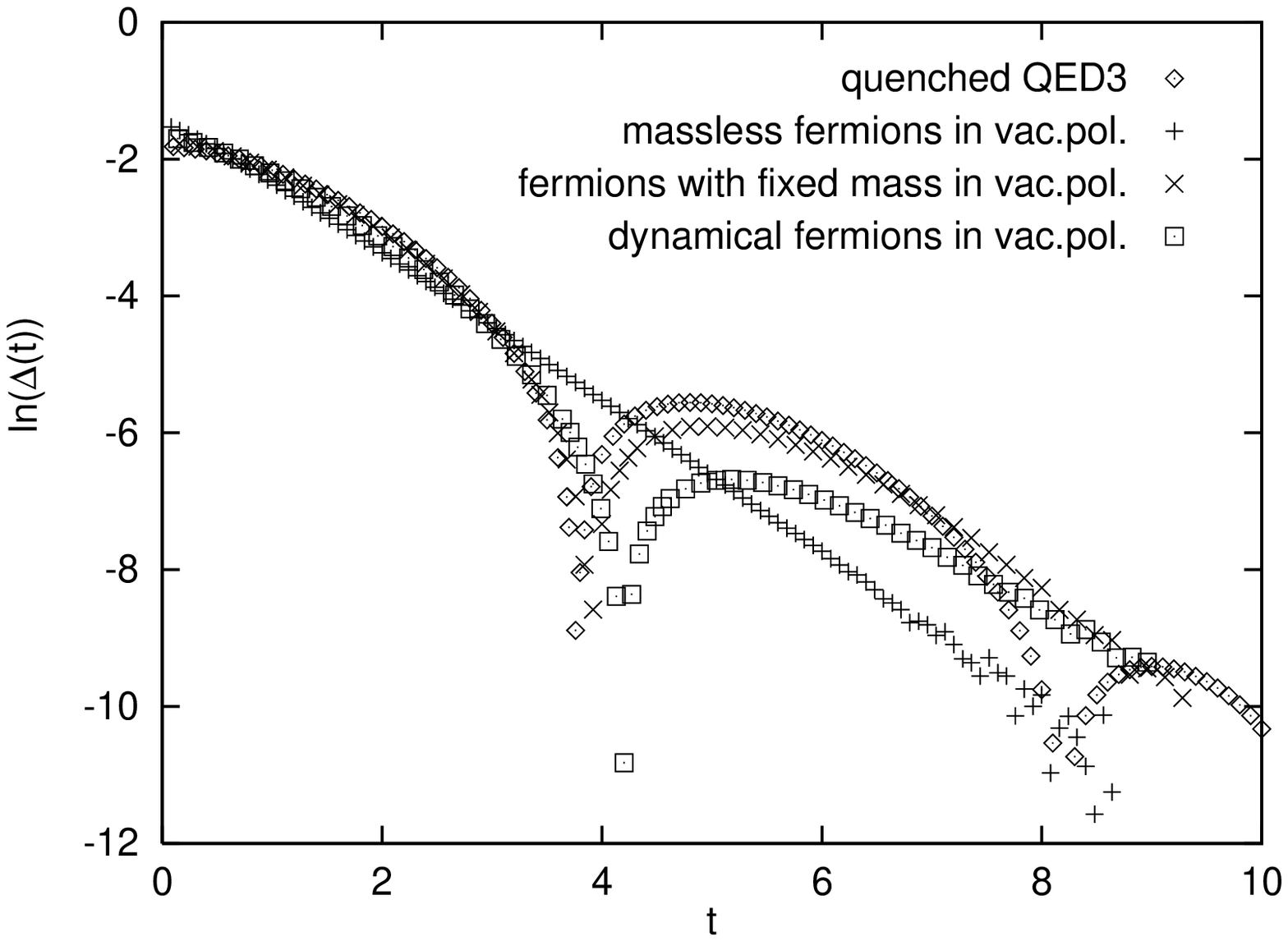,height=5.1cm}\caption{\label{qed3fig}
$\Delta(T):= -\sigma_S(T)$ from Eq.~(\protect\ref{sigmasT}) for QED$_3$ with
$2$ flavours of fermion.  ``Diamond:'' confining, quenched theory; ``plus,''
massless fermions used to evaluate the photon vacuum polarisation tensor;
``cross,'' as before but with fixed-mass fermions; ``box,'' same again but
fermions with a momentum-dependent mass function.  (Adapted from
Ref.~\protect\cite{pieterQED3}.)}}

In Ref.~\cite{pieterQED3} the QED$_3$ gap equation is solved for all four
cases and the fermion propagator analysed.  The results are summarised by
Fig.~\ref{qed3fig}.  In the quenched theory, Eq.~(\ref{Vrquench}), the
dressed-fermion $2$-point function exhibits exactly those periodic
singularities that, via Eq.~(\ref{oscexamp}), are indicative of complex
conjugate poles.  Hence this feature of the $2$-point function, tied to the
violation of reflection positivity, is a clear signal of confinement in the
theory.  That is emphasised further by a comparison with the theory that is
unquenched via massless fermions in the vacuum polarisation,
Eq.~(\ref{Vrunquench0}).  As I have described, that theory is not confining
and in this case $\sigma_S(T)$ has the noninteracting, unconfined free
particle form in Eq.~(\ref{sigmasT}).  The difference could not be more
stark.  The remaining two cases exhibit the periodic singularities that
signal confinement, just as they should based on Eq.~(\ref{Vrqed3}).

At this point I note that any concern that the presence of complex conjugate
singularities in coloured $n$-point functions leads to a violation of
causality is misguided.  Microscopic causality only constrains the
commutativity of operators, and products thereof, that represent elements in
the space of observable particle states; i.e., the space spanned by
eigenstates of the Hamiltonian.  Since Schwinger functions that violate
reflection positivity do not have a continuation into that space there can be
no question of violating causality.  It is only required that ${\cal
S}$-matrix elements that describe colour-singlet to colour-singlet
transitions should satisfy the axioms, including reflection positivity.

The violation of reflection positivity by\linebreak coloured $n$-point
functions is a sufficient condition for confinement.  However, it is not
necessary, as the example of planar, two-dimensional QCD
shows~\cite{einhorn}.  There the fermion two-point function exhibits
particle-like singularities but\linebreak the colour singlet meson bound
state amplitudes, obtained from a Bethe-Salpeter equation, vanish at momenta
coincident with the constituent-fermion mass shell.  This excludes the pinch
singularities that would otherwise lead to bound state break-up and
liberation of the constituents.  It is a realisation of confinement via a
failure of the cluster decomposition property [CDP]~\cite{gj81,sw80}.

The CDP is a requirement that the difference between the vacuum expectation
value of a product of fields and all products of vacuum expectation values of
subsets of these fields must vanish faster than any power.  [This is modified
slightly in theories, like QED, with a massless, asymptotic state: the photon
in this case.]  It can be understood as a statement about charge screening
and its failure means that, irrespective of the separation between sources,
the interaction between them is never negligible.  That is an appealing,
intuitive representation of confinement.  Failure of the CDP is an implicit
basis for confinement in the bulk of QCD potential models; e.g., Refs.\
\cite{ribeiro}.

\section{Gap Equation's Kernel}
\label{sec.kernel}
Strong interaction phenomena are characterised by DCSB and colour
confinement.  At low energy, DCSB is the more important; for example, in its
absence the $\pi$- and $\rho$-mesons would be nearly degenerate and at the
simplest observational level that would lead to a markedly different line of
nuclear stability.  These phenomena can be related to the infrared behaviour
of elementary Schwinger functions in QCD and in this section I elucidate some
constraints they place on this behaviour.

In Eq.~(\ref{ouransatz}) I described an {\it Ansatz}$\,$ for the kernel in
the quark DSE and used it to elaborate on the phenomenon of DCSB, arguing
that a good description of light-meson observables {\it requires} the kernel
to exhibit a significant enhancement in the infrared, Eq.(\ref{Dvalue}).  An
obvious question is: ``How far-reaching is this result?''

In general, as is clear from Eq.~(\ref{ouransatz}), the kernel is a product
of two terms: the dressed-gluon propagator and the dressed-quark-gluon
vertex.  For $k^2\geq 1$--$2\,$GeV$^2$ a perturbative analysis for this
product is reliable and Eq.~(\ref{ouransatz}) becomes an identity with ${\cal
G}(k^2) \to 4\pi\alpha(k^2)$.  This means that any model-dependence in the
{\it Ansatz} is constrained to the infrared domain: $k^2< 1$--$2\,$GeV$^2$.

In contemporary DSE applications to QCD it is common to build {\it
Ans\"atze}$\,$ for the higher $n$-point dressed-Schwinger functions and
employ them in developing intuition about the simpler functions.
Section~\ref{sec.dcsb} provides an illustration.  In pursuing this certain
constraints must be obeyed.

\label{discuss.vertex}
Of particular interest here is the dressed-\linebreak quark-glu\-on vertex,
which is a fully-amputated $3$-point Schwinger function.  It satisfies an
integral equation that takes the form of an inhomogeneous Bethe-Salpeter
equation.  The kernel involves $K$: the fully-amputated, quark-antiquark
scattering kernel, which by definition does not contain quark-antiquark to
single gauge-boson annihilation diagrams, such as would describe the leptonic
decay of the pion, nor diagrams that become disconnected by cutting one quark
and one antiquark line.  It also involves the scattering kernels for:
$q$-$\bar q$ to 2-gluon, $K^{2g}$; $q$-$\bar q$ to ghost-antighost,
$K^{gh\bar{gh}}$; and $q$-$\bar q$ to 3-gluon, $K^{3g}$, and also by
definition none of these can contain single-gluon intermediate states.
Hence, just as in the chiral limit a massless pole in the pseudovector vertex
signals the presence of a massless [pion] bound state, a massless,
particle-like singularity [see Eq.~(\ref{plsing})] in $\Gamma_\nu(q,p)$
signals the presence of a colour-octet bound state in the scattering
matrices: $M:= K/[ 1 - (SS)K]$; $M^{2g}:= K^{2g}/[ 1 - (DD)K^{2g}]$; etc.  As
no such coloured bound states has been observed, one must be sceptical of
calculations or {\it Ans\"atze} for any of the Schwinger functions that
entail a par\-ti\-cle-like singularity in this vertex.  [NB.\ It is
internally inconsistent to interpret as confined a gluon whose $2$-point
function violates reflection positivity whilst simultaneously asserting that
a par\-ti\-cle-like singularity in a coloured irreducible $3$-point function
does not describe an asymptotic state.]

The same objection applies to particle-like singularities in the
fully-amputated, dressed-$3$-gluon vertex, and all like coloured $n$-point
functions.  This anticipates the result of an estimate \cite{gluonv} of the
$3$-gluon vertex via a numerical simulation of lattice-QCD, which shows no
evidence for a singularity of any kind.

Rejecting particle-like singularities in\linebreak $\Gamma_\mu(q,p)$, the
possibility of an enhancement in the kernel of the gap equation can be
discussed solely in terms of the behaviour of the dressed-gluon propagator,
which in Landau gauge can be written [cf.\ Eq.\ (\ref{gluoncovariant})]
\begin{eqnarray}
D_{\mu\nu}(k) & =&  \left(\delta_{\mu\nu} - \frac{k_\mu
                k_\nu}{k^2}\right)\Delta(k^2) \,,\\
        \Delta(k^2) &:= & \frac{1}{k^2}\,{\cal P}(k^2)\,.
\end{eqnarray}
The question I posed at the beginning of this section can now be rephrased
as: ``Do observable strong interaction phenomena {\em necessarily} require
\begin{equation}
\label{irenhanced}
{\cal P}(k^2) \gg 1\; \mbox{for} \; 1 \lsim k^2/\Lambda_{\rm QCD}^2 \lsim
10\, \mbox{?''}
\end{equation}  
[$\Lambda^{N_f=4}_{\rm QCD} = 234\,$MeV.]  NB.\ Herein I do not canvass the
possibility that an irreducible vertex has a non-particle-like singularity;
i.e., a singularity of the form $(k^2)^{-\alpha}$, $\alpha>1$.  While that
evades the constraint I have elucidated, there is no indication of such
behaviour in any study to date.

I do not have an answer to the question in Eq.~(\ref{irenhanced}) but the
alternatives can be explored.  The antithesis is the extreme possibility that
\begin{equation}
\label{hype}
{\cal P}(k^2=0)=0\,,\;{\cal P}(k^2)\leq 1\;\forall\, k^2\,,
\end{equation}
which was canvassed in Refs.\ \cite{stingl123}.  [``Extreme'' because it
corresponds to a screening of the fer\-mi\-on-fer\-mi\-on interaction, as
familiar in an electrodynamical plasma, rather than the antiscreening often
discussed in zero-temperature chromodynamics.]

\FIGURE[ht]{\epsfig{figure=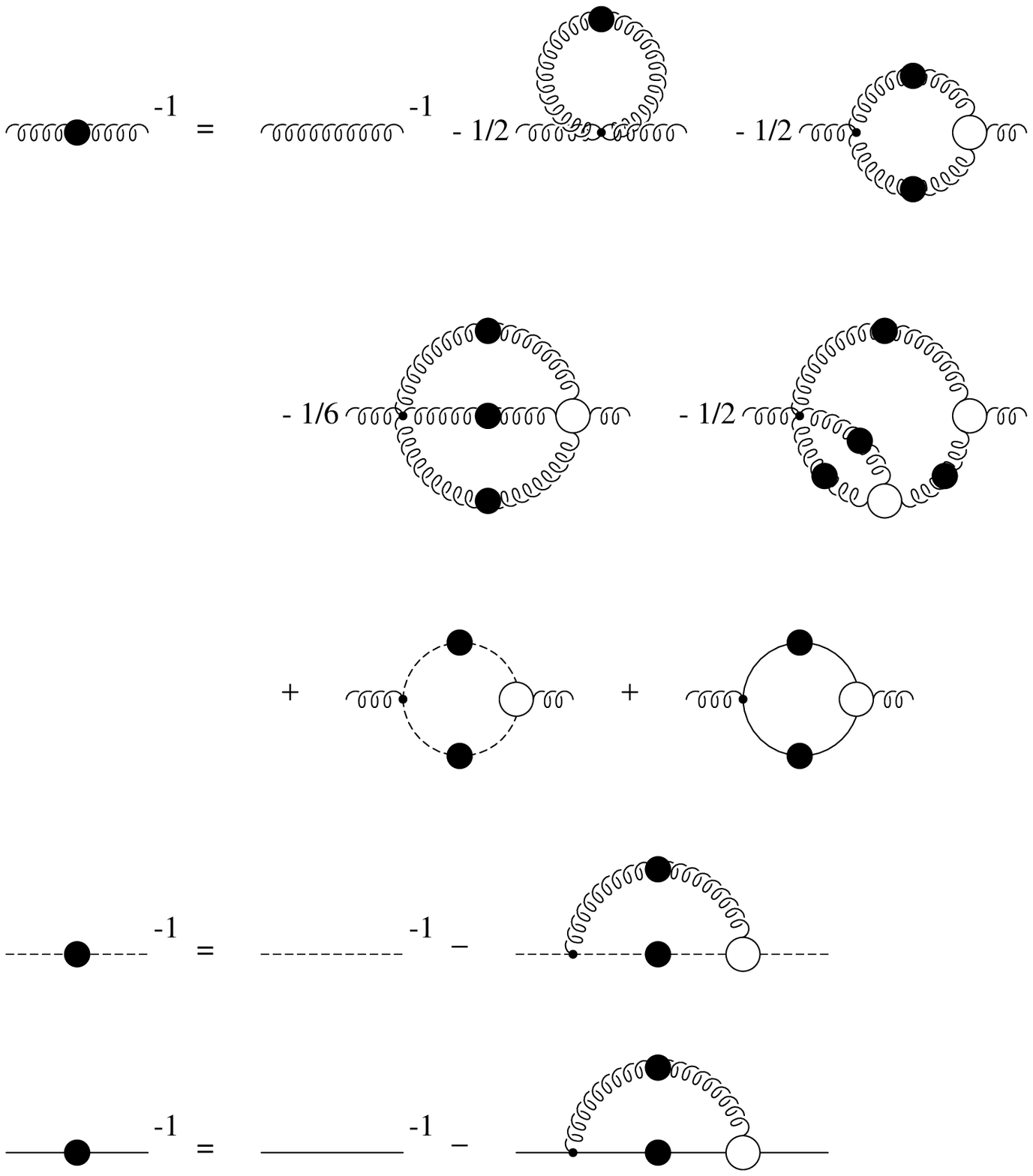,height=8.3cm}%
\caption{\label{gluonDSE} From top to bottom, depictions of the DSEs for the
gluon (spring), ghost (dashed-line) and quark (solid-line) $2$-point
functions.  Following convention, a filled circle denotes a fully dressed
propagator and an open circle, a one-particle irreducible vertex; e.g., the
open circle in the first line represents the dressed-three-gluon vertex.  The
figure illustrates the interrelation between elements in the tower of DSEs:
the gluon propagator appears in the DSE for the quark and ghost propagator;
the ghost and quark propagator in the DSE for the gluon, etc.  (Adapted from
Ref.~\protect\cite{hauck}.)}}

References$\,$\cite{stingl123} \label{firststingl}proposed solving the
DSEs\linebreak via rational polynomial {\it Ans\"atze} for the
one-par\-ti\-cle irreducible components of the Schwinger\linebreak functions.
This method attempts to preserve aspects of the organising principle of
perturbation theory in truncating the DSEs.  In connection with the gluon
DSE, depicted in Fig.~\ref{gluonDSE}, these studies employed a truncation
that, for simplicity: retains only the first, third and sixth diagrams on the
r.h.s.\ of the first equation in Fig.~\ref{gluonDSE}; neglects the last
[fermion] equation; and employs the leading order solution of the ghost
equation, which has the appearance of the massless free propagator: $\sim
1/k^2$.  In this approach the\linebreak {\it Ans\"atze} for the $3$-gluon and
quark-gluon vertices exhibit {\em ideal} particle-like poles [$\alpha = 1$ in
Eq.\ (\ref{plsing})].  Since these poles are an essential element of the
solution procedure then, in the absence of a physically sensible
interpretation or explanation of them, one could simply reject this result.

Alternatively, one can suppose that Eq.~(\ref{hype}) is more robust than the
procedure employed to motivate it and explore the phenomenological
consequences of the conjecture \cite{stingl123}:
\begin{equation}
\label{stinglprop}
{\cal P}_S(k^2):= k^4/(k^4+b^4)\,, 
\end{equation}
where $b$ is a dynamically generated mass scale.  Following this approach it
was found that if there are no particle-like singularities in the
quark-glu\-on vertex, $\Gamma_\nu(q,p)$, then ${\cal P}_S(k^2)$ is unable to
confine quarks~\cite{fredIR,axelIR} and $b$ must be fine-tuned to very small
values:
\begin{equation}
b< b_c \simeq \Lambda_{\rm QCD}\,,
\end{equation}
if DCSB is to occur$\,$\cite{fredIR,papa1,axelIR,natale,fredIRnew}.  It is
therefore apparent that Eq.~(\ref{hype}) is phenomenologically difficult to
maintain.  [NB.\ Achieving DCSB by requiring $b \sim 0$ is indicative of the
{\it dynamical evasion} of Eq.~(\protect\ref{hype}) since ${\cal P}_S(k^2)
\to 1$ rapidly for small values of $b$.]

Nevertheless, the hypothesis has been explor\-ed in studies~\cite{marenzoni}
of the dressed-gluon $2$-point function using numerical simulations of
lattice-QCD.  ${\cal P}(k^2=0)$ is necessarily finite in simulations on a
finite lattice because of the inherent infrared cutoff.  Thus one can only
truly determine ${\cal P}(k^2 \sim 0)$ by considering the behaviour of the
numerical result in both the countable limit of infinitely many lattice sites
and the continuum limit.  

The form ${\cal P}_S(k^2)$ does not provide as good a fit to the lattice data
as an alternative form, which in the countable limit is
\begin{equation}
\label{Plat}
{\cal P}_L(k^2) := \frac{k^2}{M^2 + Z \,k^2
        \,\left( k^2 a^2 \right)^\eta}\,,        
\end{equation}
$ 0 < k^2 < 0.6/a^2 \, \sim 50 \,\Lambda_{\rm QCD}^2$, where $1/a \approx
2.0\,$GeV is the inverse lattice spacing, $Z\approx 0.1$, $\eta \approx
0.53$, and $M\approx 0.16\,$GeV.  This takes the maximum value 
\begin{equation}
{\cal P}_L(k^2 = 21 \Lambda_{\rm QCD}^2) = 13.6
\end{equation}
and corresponds to a less extreme alternative to Eq.\ (\ref{hype}), which I
shall characterise as
\begin{equation}
\label{hypeB}
{\cal P}(k^2=0)=0\,,\;{\rm max}\left({\cal P}(k^2)\right) \lsim {\cal O}(10)\,.
\end{equation}
The feature ${\cal P}(k^2=0)=0$ is critically dependent on whether or not $M$
is nonzero.  It appears to be nonzero in the countable limit but, as
emphasised in Ref.~\cite{marenzoni}, the behaviour of $M$ (and $\eta$) in the
continuum limit is unknown.  [NB.\ All existing lattice-QCD simulations of
the gluon propagator; e.g., Ref.~\cite{skull}, yield fitted forms that lie in
the class specified by Eq.~(\ref{hypeB}).  A dressed-gluon propagator
satisfying Eq.~(\ref{hype}) automatically satisfies Eq.~(\ref{hypeB}).]

The phenomenological implications of Eq.\linebreak (\ref{Plat}) can be
explored using the methods of Ref.\ \cite{fredIR}.  A preliminary estimate
follows from observing that ${\cal P}_L(k^2)$ is approximately equivalent to
${\cal P}_S(k^2)$ if one identifies
\begin{equation}
\label{blattice}
b_L \sim \sqrt{M/a} = 0.57\,{\rm GeV}.  
\end{equation}
Hence one expects that Eq.\ (\ref{Plat}) does not generate DCSB nor confine
quarks.  [A value of $b\approx 0.4\,$GeV$>b_c$ in ${\cal P}_S(k^2)$ provides
the best fit to the lattice data and this supports the same
conclusion.\label{Stingl-like}] In order to quantitatively verify this
conclusion I note that: it is the combination
\begin{equation}
g^2{\cal P}(k^2)/k^2
\end{equation}
that appears in Eq.\ (\ref{gendse}) and $g^2$ is not determined in Ref.\
\cite{marenzoni}; and one must extrapolate ${\cal P}_L(k^2)$ outside the
fitted domain.  Both of these requirements are fulfilled if: 1) one assumes
that a one-loop perturbative analysis is reliable for $k^2 \gsim 25
\,\Lambda_{\rm QCD}^2$; and 2) employs
\begin{equation}
\label{pourl}
g^2\,{\cal P}_l(k^2) := 
\left\{
\begin{array}{ll}
g_m^2 {\cal P}_L(k^2)\,,&  k^2\leq k_m^2\\
g^2(k^2)\,,               & k^2> k_m^2\,,
\end{array}
\right.
\end{equation}
with $g^2(k^2)$ the one-loop running coupling, requiring that
$\Delta_l(k^2):= {\cal P}_l(k^2)/k^2$ and its first derivative be continuous
at $k_m^2$.  This procedure yields $\Delta_l(k^2)$ in Fig.$\,$\ref{figgprop}
with
\begin{equation}
g_m = 0.65\,,\; k_m^2 = 30\,\Lambda_{\rm QCD}^2\,.
\end{equation}
\FIGURE[ht]{\epsfig{figure=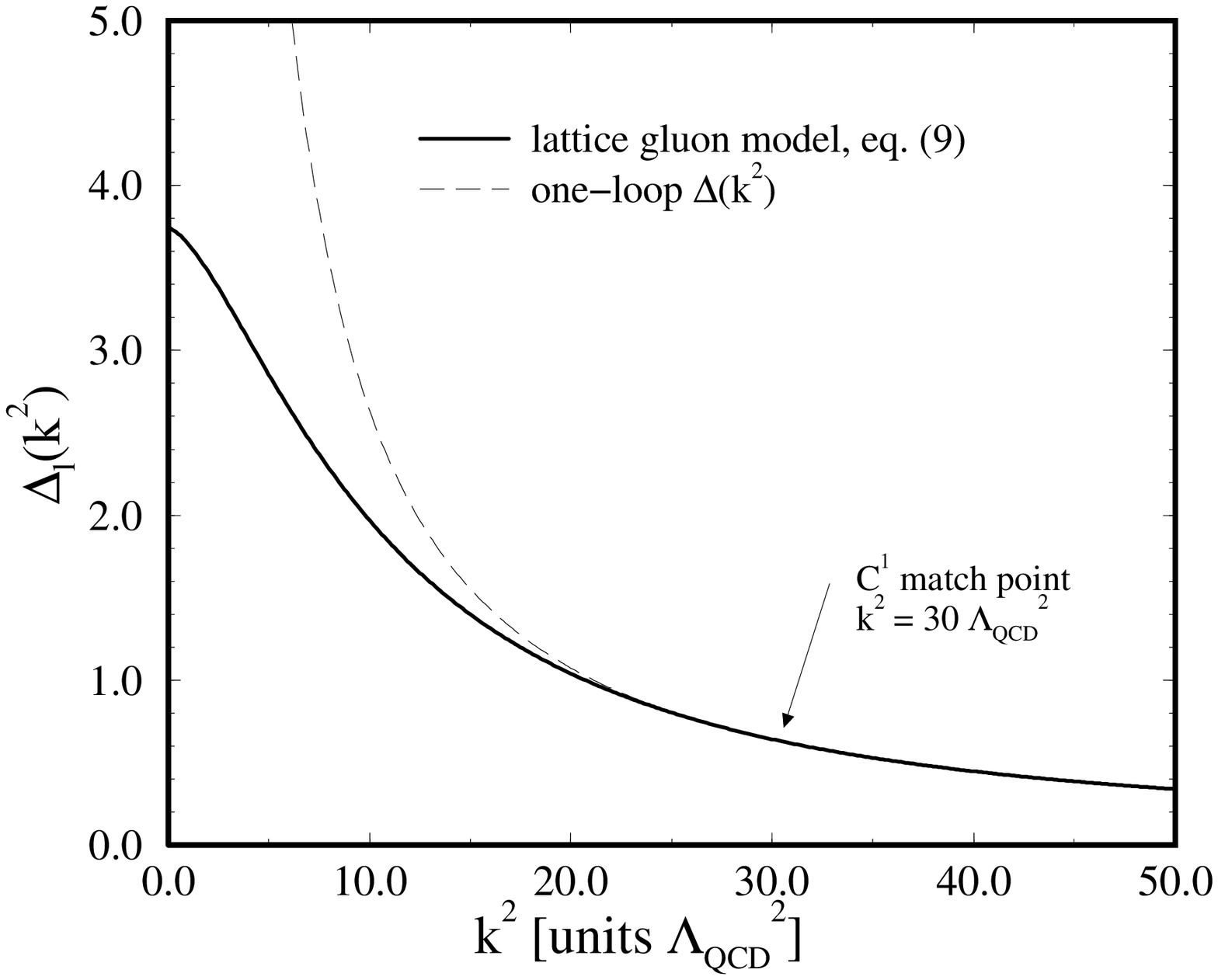,height=5.9cm}\caption{$\Delta_l(k^2):=
{\cal P}_l(k^2)/k^2$ from Eq.\ (\protect\ref{pourl}). ${\cal P}_l(k^2)$ is
Eq.\ (\protect\ref{Plat}) in the infrared and extrapolates this lattice model
outside the domain accessible in the
simulation~\protect\cite{marenzoni,skull}.  (Adapted from Ref.\
\protect\cite{fredIRnew}.)\label{figgprop}}}

It is now straightforward to solve Eq.\ (\ref{gendse}) with a variety of {\it
Ans\"atze} for the quark-gluon vertex that do not exhibit particle-like
singularities. This was pursued in Ref.\ \cite{fredIRnew}, using the methods
described in Sec.\ \ref{sec.quark} and renormalising at the momentum cutoff,
$\Lambda_{\rm UV} \sim 10^4 \Lambda_{\rm QCD}$, for simplicity, since the
$p^2$-evolution of $A(p^2)$ and $B(p^2)$ beyond that point is completely
determined by $g^2(k^2)$.  That study employed the bare vertex
$\Gamma_\mu(p,q):= \gamma_\mu$; the {\it Ansatz}$\,$\cite{bc80}:
\begin{eqnarray}
\label{bcvtx}
\lefteqn{i\Gamma^{\rm BC}_\nu(p,q):= i\Sigma_A(p^2,q^2)\,\gamma_\nu +
(p+q)_\nu\,}\\ && \nonumber \times\left[\sfrac{1}{2}i\gamma\cdot (p+q) \,
\Delta_A(p^2,q^2) + \Delta_B(p^2,q^2)\right]\,,
\end{eqnarray}
where 
\begin{equation}
\begin{array}{l}
\Sigma_A(p^2,q^2):= [A(p^2)+A(q^2)]/2\,, \\
\Delta_A(p^2,q^2):= [A(p^2)-A(q^2)]/[p^2-q^2]\,,\\
\Delta_B(p^2,q^2):= [B(p^2)-B(q^2)]/[p^2-q^2]\,;
\end{array}
\end{equation}
and an augmented form~\cite{cp90}
\begin{equation}
\label{cpvtx}
\Gamma^{\rm CP}_\mu(p;q)  :=  \Gamma^{\rm BC}_\mu(p,q) + 
        \Gamma^{6}_\mu(p,q)\,,
\end{equation}
\begin{eqnarray}
\lefteqn{\Gamma^{6}_\mu(p,q)  := \left[\gamma_\mu (p^2 - q^2) \right.}\\
&& \left. - (p+q)_\mu (\gamma\cdot p - \gamma\cdot
 q)\right] \frac{ A(p^2) - A(q^2)}{2 d(p,q)}\nonumber \,,
\end{eqnarray}
with 
\begin{equation}
d(p,q):= \frac{[p^2-q^2]^2 + [M(p^2)^2 + M(q^2)^2]^2}{p^2+q^2}\,,
\end{equation}
each of which allows the quark DSE to be solved in isolation; i.e., without
coupling to other DSEs.  In all cases the lattice result, Eq.\ (\ref{pourl}),
yields $M(p^2)\equiv 0$ in the chiral limit; i.e., \underline{no DCSB}.

\FIGURE[ht]{\epsfig{figure=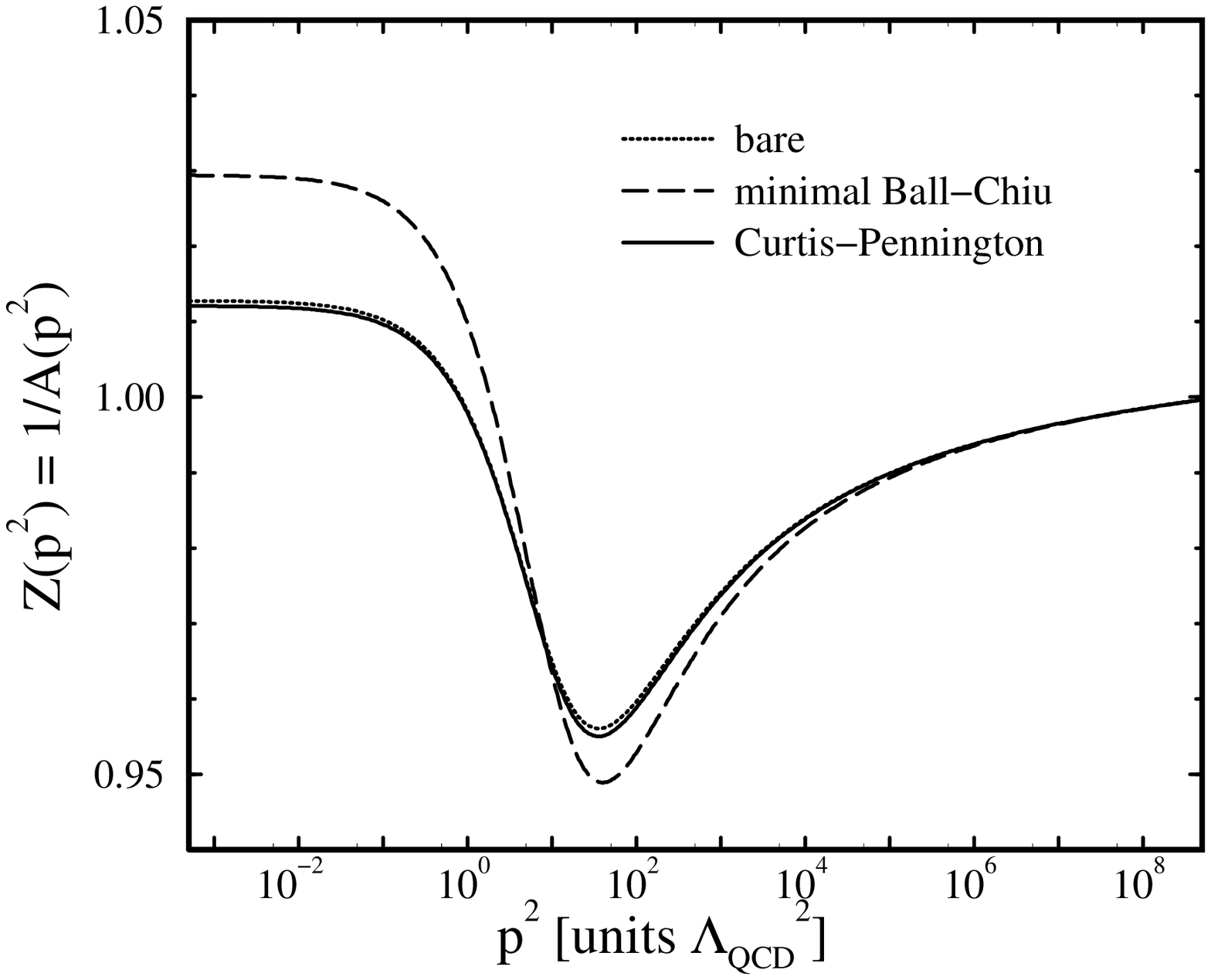,height=5.9cm}\caption{$Z(p^2)$
obtained as the solution of Eq.\ (\protect\ref{gendse}) using Eq.\
(\protect\ref{pourl}) with: Eq.\ (\protect\ref{bcvtx}) - solid line; Eq.\
(\protect\ref{cpvtx}) - dashed line; and $\Gamma_\mu(p;q)= \gamma_\mu$ -
dotted line.  That Eq.\ (\protect\ref{Plat}) does not confine quarks is
manifest in the result: $Z(p^2=0)\neq 0$, which is independent of the vertex
{\it Ansatz}.  (Adapted from Ref.\ \protect\cite{fredIRnew}.)\label{figa}}}
The absence of DCSB means it is straightforward to decide whether
(\ref{pourl}) generates confinement.  Following Sec.\ \ref{sec.conf},
confinement is manifest if, for $p^2\simeq 0$,
\begin{equation}
Z(p^2) \propto (p^2)^\alpha,\; \alpha\geq 1,
\end{equation}
in which case the dressed-quark propagator does not have a Lehmann
representation; i.e., violates the axiom of reflection positivity.  The
solution obtained with the vertex {\it Ans\"atze} introduced\linebreak above
is depicted in Fig.\ \ref{figa}: the behaviour is qualitatively equivalent in
each case and demonstrates explicitly that Eq.\ (\ref{Plat}) does not
generate confinement.

These failures: the absence of both DCSB and confinement, confirm the
preliminary hypothesis based on the correspondence with ${\cal P}_S$ via an
effective value of $b$, Eq.\ (\ref{blattice}).  [The same conclusion applies
to the result in Ref.\ \protect\cite{skull}, which is is pointwise smaller in
magnitude $(\lsim 1/3)$ than Eq.\ (\protect\ref{Plat}) on the entire fitted
domain.]

Equation~(\ref{gendse}) was also solved for $Z(p^2)$ using
\begin{equation}
\tilde{\cal P}_l(k^2):= \left( 1 
                + \varsigma\,{\rm e}^{-k^2/\Lambda_{\rm QCD}^2}\right)
                {\cal P}_l(k^2)
\end{equation}
where $\varsigma$ is a variable ``strength'' parameter.  Increasing
$\varsigma$ moves the peak in $\tilde{\cal P}_l(k^2)$ toward $k^2=0$ and
increases its height, thereby making it increasingly like the model of
Eq.~(\ref{ouransatz}).  The form of $Z(p^2)$ is qualitatively unchanged and
hence there is no signal for the onset of confinement until $\varsigma\gsim
300$.  At $\varsigma= 300$ the maximum value is
\begin{equation}
\tilde{\cal P}_l(k^2=0.98\,\Lambda_{\rm QCD}^2)= 210
\end{equation}
and 
\begin{equation}
\begin{array}{llcr}
& \tilde{\cal P}_l(0.98\,\Lambda_{\rm QCD}^2)/
        \tilde{\cal P}_l(30\,\Lambda_{\rm QCD}^2) & = & 16.0\,, \\
\mbox{cf.}\;& 
{\cal P}_l(21\,\Lambda_{\rm QCD}^2)/
        {\cal P}_l(30\,\Lambda_{\rm QCD}^2) & = & 1.0\,.
\end{array}
\end{equation}
The model of Eq.~(\ref{ouransatz}) exhibits a peak at $k^2=
3.7\,\Lambda^2_{\rm QCD}$ and the value of the ratio introduced above is
$44$, neglecting only for the purpose of this comparison the purely
long-range, $\delta^4(k)$-part of that interaction.  A comparison of 
\begin{equation}
g_m^2 \tilde{\cal P}_l(\Lambda_{\rm QCD}^2)\approx 89
\end{equation}
with the critical coupling of $g_c^2 \approx 11$ in Refs.\
\cite{higashijimaportermelbourne} shows that such large values of $\varsigma$
ensure DCSB.

To recapitulate: in this section we have seen that in the absence of
particle-like singularities in the dressed-quark-gluon vertex, extant
proposals for the dres\-sed-gluon propagator that satisfy Eq.~(\ref{hypeB})
neither confine quarks nor break chiral symmetry dynamically.  This class
includes all existing estimates of ${\cal P}(k^2)$ via numerical simulations.

\subsection{Critical Interaction Tension}
\label{sub.polarisationtension}
Furthermore the calculations described in this section reaffirm the
long-standing DSE result~\cite{higashijimaportermelbourne} that the existence
of DCSB in QCD places a constraint on the minimum phenomenologically
acceptable value of the ``interaction tension:''
\begin{eqnarray}
\label{sigmadelta}
\sigma^{\Delta} & := & \frac{1}{4\pi}\int_{\Lambda_{\rm QCD}^2}^{\Lambda_{\rm
pQCD}^2}\,dk^2\,k^2\,\Theta(k^2)\,,\\
\Theta(k^2) & := & g^2\Delta(k^2) - g^2\Delta(\Lambda_{\rm pQCD}^2) \,.
\label{theta}
\end{eqnarray}
The lower cutoff is $\Lambda_{\rm QCD}$ because string-breaking becomes
effective for lengths $\gsim 1/\Lambda_{\rm QCD}$ $\sim 1\,$fm, while the
upper cutoff, $\Lambda_{\rm pQCD}=2\,$GeV, marks the boundary above which the
interaction is calculable in perturbation theory.  [NB.\ If a mechanism can
be found by which singularities in the dressed-quark-gluon vertex become
phenomenologically tenable then this constraint persists with the only change
being $g^2 \Delta(k^2)$ $\to$ ${\cal G}(k^2)/k^2$; i.e., the effective
interaction.]

The lattice result, Eq.~(\ref{pourl}), yields 
\begin{equation}
\sigma^\Delta_l \simeq 0.35\, {\rm GeV}^2 \sim 6\,\Lambda^2_{\rm QCD},
\end{equation}
while from Ref.\ \cite{fredIR} I estimate the critical value for DCSB to be:
\begin{equation}
\sigma^{\Delta} > \sigma^{\Delta}_c \sim 0.5\, {\rm GeV}^2 \sim
9\,\Lambda^2_{\rm QCD}\,,
\end{equation}
with the exact value depending a little [$\sim 20$\%] on model details.  This
is the critical value, which only supports an incipient quark condensate;
i.e., $\langle \bar q q \rangle$ infinitesimally-greater-than zero.  For
agreement with its phenomenological value one re\-qui\-res \cite{pieterVM}:
\begin{equation}
\sigma_{\rm phen.}^\Delta \sim 4\,{\rm GeV}^2\sim 70\,\Lambda^2_{\rm QCD};
\end{equation}
i.e., on order-of-magnitude larger.

One additional observation: following Refs. \cite{gastao,furnstahl}, an
estimate of the gluon condensate can be obtained from
\begin{equation}
\label{condensate}
\langle \alpha G G \rangle := 
\frac{3}{4\pi^3}\int_{\Lambda_{\rm QCD}^2}^{\Lambda_{\rm
pQCD}^2}\,dk^2\,k^4\,\Theta(k^2)\,.
\end{equation}
The lattice model, Eq.~(\ref{pourl}), yields $0.09\,$GeV$^4$; the critical
value for DCSB is $\approx$ $0.12$ GeV$^4$; and the model of Ref.\
\cite{pieterVM} yields $0.42\,$GeV$^4$.  Contemporary analyses of QCD Sum
Rules employ a value~\cite{derek}: $0.095\pm0.05\,$GeV$^4$.  However, this
cannot be directly compared with the calculated values reported here because
of the ambiguity in subtracting the ``perturbative contribution'' to the
gluon correlator.  Equations (\ref{sigmadelta}) - (\ref{condensate}), with a
subtraction that is constant below $k^2 = \Lambda_{\rm pQCD}^2$, employ just
one of many possibilities.  Nevertheless the calculated results indicate the
relevant scales.

\section{Gluon DSE}
\label{sec.gluon}
Plainly now it is a model-independent result that DCSB requires a significant
infrared enhancement in the kernel of QCD's gap equation.  Such an
enhancement can also yield confinement, by ensuring that coloured $n$-point
Schwinger functions violate reflection positivity, as discussed in
Sec. \ref{sec.conf}.  The question that naturally arises is: ``Where does
this enhancement come from?''  Some guidance may be sought in studies of the
DSE satisfied by the dressed-gluon propagator, which is depicted in
Fig.~\ref{gluonDSE}.  However, as I now describe, these studies are
inconclusive.

Early analyses~\cite{bbz} used the ghost-free axial gauge: $n\cdot A^a = 0$,
$n^2>1$, in which case the second equation in Fig.\ \ref{gluonDSE} is absent
and two independent scalar functions: $F_1$, $F_2$, are required to fully
specify the dressed-gluon propagator, cf.\ the covariant gauge expression in
Eq.~(\ref{gluoncovariant}), which requires only one function.  In the absence
of interactions: $F_1(k^2) = -1/k^2$, $F_2(k^2)\equiv 0$.  These studies
employed an {\it Ansatz} for the three-gluon vertex that doesn't possess a
particle-like singularity and neglected the coupling to the quark DSE.  They
also assumed $F_2\equiv 0$, even nonperturbatively, and ignored it in solving
the DSE.  The analysis then yielded
\begin{equation}
\label{enhanced}
F_1(k^2) \stackrel{k^2\to 0}{\sim} \frac{1}{k^4}\,;
\end{equation}
i.e., a marked infrared enhancement that can yield an area law for the Wilson
loop~\cite{west82} and hence confinement, and DCSB as described above, {\it
without} fine-tuning.

This effect is driven by the gluon vacuum polarisation, diagram three in the
first line of Fig.~\ref{gluonDSE}.  A similar result was obtained in
Ref.~\cite{atkinson83}.  However, a possible flaw in these analyses was
identified in Ref.~\cite{west83}, which argued from properties of the
spectral density in ghost-free gauges that $F_2$ cannot be zero but acts to
cancel the enhancement in $F_1$.  [NB.\ Retaining $F_2$ in the analysis
yields a coupled system of equations for the gluon propagator that is at
least as complicated as that obtained in covariant gauges, which perhaps
outweighs the apparent benefit of eliminating ghost fields in the first
place.]

There have also been analyses of the gluon DSE using Landau gauge and those
of Refs.\linebreak \cite{mandelstam,brown89} are unanimous in arriving at the
covariant gauge analogue of Eq.~(\ref{enhanced}), again driven by the gluon
vacuum polarisation diagram.  In these studies {\it Ans\"atze} were used for
the dressed-three-gluon vertex, all of which were free of particle-like
singularities.  However, these studies too have weaknesses: based on an
anticipated dominance of the gluon-vacuum polarisation, truncations\linebreak
were implemented so that only the third and fifth diagrams on the r.h.s.\ of
the first equation in Fig.~\ref{gluonDSE} were retained.  In covariant gauges
there is {\it a priori} no reason to neglect the ghost loop contribution,
diagram six, although per\-tur\-ba\-tive\-ly its contribution was estimated
to be small~\cite{brown89}.

As described on page~\pageref{firststingl}, the Landau gauge studies of
Refs. \cite{stingl123} yield a qualitatively different result: Eq.\
(\ref{stinglprop}), but the question of how the particle-like singularities
in the associated vertex {\it Ans\"atze} can be made consistent with the
absence of coloured bound states in the strong interaction spectrum is
currently unanswered.  Nonetheless proponents of the result in
Eq.~(\ref{stinglprop}) claim support from
studies~\cite{zwanzigerold,zwanziger} of ``complete'' gauge fixing; i.e., in
the outcome of attempts to construct a Fadde$^\prime$ev-Popov-like
determinant that eliminates Gribov copies or ensures that the functional
integration domain for the gauge field is restricted to a subspace without
them.  Fixing a so-called ``minimal Landau gauge,'' which enforces a
constraint of integrating only over gauge field configurations inside the
Gribov horizon; i.e., on the simplest domain for which the Fadde\'ev-Popov
operator is invertible, the dressed-gluon $2$-point function is shown to
vanish at $k^2=0$.  However, the approach advocated in Refs.~\cite{stingl123}
makes no use of the additional ghost-like fields necessary to restrict the
integration domain.  [NB.\ Hitherto the quantitative effect of Gribov copies
in nonperturbative studies remains unknown.]

Recently the direct approach to solving the Landau gauge gluon DSE, pioneered
in Refs.\linebreak \cite{mandelstam,brown89}, has been revived by two groups:
${\cal A}$, Refs. \cite{hauck,hauckPRL,hauckAdelaide}; and ${\cal B}$,
Refs.~\cite{bloch,bloch2,davidAdelaide,jacquesAdelaide}, with the significant
new feature that nonperturbative effects in the ghost sector are admitted;
i.e., a nonperturbative solution of the DSE for the ghost propagator is
sought in the form
\begin{equation}
G^{ab}(k) = - \delta^{ab}\,\frac{\varpi(k^2)}{k^2}\,.
\end{equation}
[Without interactions, $\varpi(k^2) \equiv 1$.]

These studies analyse a truncated gluon-\linebreak ghost DSE system,
retaining only the third and sixth loop diagrams in the first equation of
Fig.~\ref{gluonDSE}, and also the second equation.  Superficially this is the
same complex of equations as studied in Refs.~\cite{stingl123}.  However, the
procedure for solving it is different, arguably less systematic but also less
restrictive.  

The difference between the groups is that ${\cal A}$ employ {\it Ans\"atze}
for the dressed-ghost-gluon and dressed-three-gluon vertices con\-struc\-ted
so as to satisfy the relevant Slavnov-Taylor identities while ${\cal B}$
simply use the bare vertices.  Nevertheless they agree in the conclusion that
in this truncation the infrared behaviour of the gluon DSE's solution is
determined by the ghost loop alone: it overwhelms the gluon vacuum
polarisation contribution.  

That is emphasised in Ref.~\cite{davidAdelaide}, which eliminates every loop
diagram in truncating the first equation of Fig.~\ref{gluonDSE} {\it except}
the ghost loop and still recovers the behaviour of
Ref.~\cite{jacquesAdelaide}.  That behaviour is
\begin{equation}
\label{irAB}
\varpi(k^2) \sim \frac{1}{(k^2)^{\kappa}} ,\;
d(k^2)\sim (k^2)^{2\kappa}\,,
\end{equation}
for $k^2\lsim \Lambda^2_{\rm QCD}$, with $0.8\lsim \kappa \leq 1$.  Exact
evaluation of the angular integrals that arise when solving the integral
equations gives the integer-valued upper bound, $\kappa = 1$~\cite{bloch2}.
This corresponds to a dressed-gluon $2$-point function that vanishes at
$k^2=0$, although the suppression is very sudden with the propagator not
peaking until $k^2\approx\Lambda^2_{\rm QCD}$, where
\begin{equation}
(\left.d(k^2)/k^2)\right|_{k^2=\Lambda_{\rm QCD}^2} \sim 100/\Lambda_{\rm
QCD}^2\,; 
\end{equation}
i.e., it is very much enhanced over the free propagator.  [See, e.g.,
Ref.~\cite{hauck}, Fig.~12.]  $\kappa = 1$ also yields a dressed-ghost
propagator that exhibits a dipole enhancement analogous to that of
Eq.~(\ref{enhanced}).  

A [renormalisation group invariant] running strong-coupling consistent with
the truncations that yield these solutions is:
\begin{equation}
\alpha(k^2) := \sfrac{1}{4\pi}\,g^2\,\varpi^2(k^2)\,d(k^2)
\end{equation}
and its value at $k^2=0$ is fixed by the numerical solutions:
\begin{equation}
\label{runningcouplings}
\begin{array}{l|c|c}
                & {\cal A} & {\cal B} \\\hline
\alpha(k^2=0)   &   9.5       & \sim 4\;\mbox{or}\;12
\end{array}\,.
\end{equation}
[NB.\ Group ${\cal A}$ approximates the angular integrals and uses vertex
{\it Ans\"atze}.  Group ${\cal B}$ uses bare vertices and in
Ref.~\cite{bloch} approximates the angular integrals to obtain $\alpha(0)\sim
12$, while in Ref.~\cite{bloch2} the integrals are evaluated exactly, which
yields $\alpha(0)=\sfrac{4}{3}\pi\approx 4.2$.]

The qualitative feature common to both\linebreak groups is that the
Grassmannian ghost loops act to suppress the dressed-gluon propagator in the
infrared.  That may also be said of Refs.~\cite{stingl123}.  [Indications
that the quark loop, diagram seven in Fig.~\ref{gluonDSE}, acts to oppose an
enhancement of the type in Eq.~(\ref{enhanced}) may here, with hindsight, be
viewed as suggestive.]  

One aspect of ghost fields is that they enter because of gauge fixing via the
Fadde$^\prime$ev-Popov determinant.  Hence, while none of the groups
introduce the additional Fadde$^\prime$ev-Popov contributions advocated in
Refs.~\cite{zwanzigerold,zwanziger}, they nevertheless do admit ghost
contributions, and in their solution the number of ghost fields does not have
a qualitative impact.  Reference~\cite{zwanziger} also obtains a
dressed-propagator for the Fadde$^\prime$ev-Popov fields with a $k^2=0$
dipole singularity.  It contributes to the action via the term employed to
restrict the gauge field integration domain, in which capacity the dipole
singularity can plausibly drive an area law for Wilson loops.

Schwinger functions are the primary object of study in numerical simulations
of lattice-QCD and Refs.~\cite{marenzoni,skull} report contemporary estimates
of the lattice Landau gauge dressed-gluon $2$-point function.  As we saw in
Sec. \ref{sec.kernel}, they are consistent with a finite although not
necessarily vanishing value of $d(k^2 = 0)$.  However, simulations of the
dressed-ghost $2$-point function find no evidence of a dipole singularity,
with the ghost propagator behaving as if $\varpi(k^2) = 1$ in the smallest
momentum bins~\cite{schilling}.  [NB.\ Since the quantitative results from
groups ${\cal A}$ and ${\cal B}$ differ and exhibit marked sensitivity to
details of the numerical analysis, any agreement between the DSE results for
$\varpi(k^2)$ or $d(k^2)$ and the lattice data on some subdomain can be
regarded as fortuitous.]

The behaviour in Eqs.~(\ref{irAB}) also entails the presence of particle-like
singularities in extant {\it Ans\"atze} for the dressed-ghost-gluon,
dressed-\linebreak three-gluon and dressed-quark-gluon vertices that are
consistent with the relevant Slavnov-Taylor identities.  [$\kappa = 1$
corresponds to an ideal simple pole singularity.]  Hence while this behaviour
may be consistent with the confinement of elementary excitations, as
currently elucidated it also predicts the existence of coloured bound states
in the strong interaction spectrum [see page \pageref{discuss.vertex}].  

Furthermore, while it does yield a running strong-coupling with
$\alpha(k^2=0) \gsim 1$, that, as we saw in Sec. \ref{sec.kernel}, makes DCSB
dependent on fine tuning.  To make this plain, the maximum value of $12$ in
Eq. (\ref{runningcouplings}) yields $\sigma^\Delta \approx 0.7\,$GeV$^2$,
which is just above $\sigma_c^\Delta$ but falls far short of the value
required to produce the physical value of the quark condensate.  In fact, the
quark condensate is only $\sim 5$\% of the phenomenologically required value
in Eq.~(\ref{qbq1})~\cite{jacquesPrivate}.  Notwithstanding these remarks,
the studies of Refs.~\cite{hauck,hauckPRL,hauckAdelaide} and subsequently
Refs.~\cite{bloch,bloch2,davidAdelaide,jacquesAdelaide} are laudable.  They
have focused attention on a previously unsuspected qualitative sensitivity to
truncations in the gauge sector.

To recapitulate.  It is clear from Sec.\ \ref{sec.kernel} that DCSB requires
the effective interaction in the quark DSE to be strongly enhanced at $k^2
\sim \Lambda^2_{\rm QCD}$.  [Remember too that in Sec.\ \ref{sec.dcsb},
Figs.\ \ref{latZZ}, we saw that modern lattice simulations confirm the pattern
of behaviour exhibited by quark DSE solutions obtained with such an enhanced
interaction.]  Studies of QCD's gauge sector indicate that gluon-gluon and/or
gluon-ghost dynamics can generate such an enhancement.  However, the
qualitative nature of the mechanism and its strength remains unclear: is it
the gluon vacuum polarisation or that of the ghost that is the driving force?
It is a contemporary challenge to explore and understand this.

\section{Bethe-Salpeter Equation}
\label{sec.bound}
Hitherto I have focussed on the not-directly-ob\-ser\-va\-ble elementary
excitations in QCD.  They make themselves manifest in hadrons and their
properties.  In quantum field theory, two and three-body bound states are,
respectively, des\-crib\-ed by the Poincar\'e-covariant Bethe-Salpeter and
Fad\-de$^\prime$ev equations.  Solving an equation of this type yields the
bound states mass and also an amplitude that describes the bound state
constraints on the constituents' momenta.  This amplitude is a valuable
intuitive guide and, in cases where a simple quark model analogue of the
bound state exists, the amplitude incorporates and extends the information
present in that analogue's quantum mechanical wave function.

In quantum field theory, as in classical mechanics, the interacting two-body
problem is\linebreak much simpler than that of three such bodies, and I
illustrate the bound state application of DSEs by considering mesons.  The
renormalised homogeneous BSE for a bound state of a dressed-quark and
dressed-antiquark with total momentum $P$ is
\begin{equation}
\label{genbse}
[\Gamma_H(k;P)]_{tu} = \int^\Lambda_q \,
[\chi(q;P)]_{rs} K_{tu}^{rs}(q,k;P)\,,\;
\end{equation}
\begin{equation}
\chi_H(q;P):= {\cal S}(q_+)\,\Gamma_H(q;P)\,{\cal S}(q_+)\,,
\end{equation}
with: $\Gamma_H(k;P)$ the Bethe-Salpeter amplitude, where $H$ specifies the
flavour structure of the meson; ${\cal S}(p):= {\rm
diag}[S_u(p),S_d(p),S_s(p)]$; $q_+ = q+\eta_P P$, $q_-= q - (1-\eta_P) P$;
and $r,\ldots, u$ represent colour-, Dirac- and flavour-matrix indices.
[$\eta_P\in[0,1]$ is the momentum partitioning parameter.  It appears in
Poincar\'e covariant treatments because, in general, the definition of the
relative momentum is arbitrary.  Physical observables, such as the mass, must
be independent of $\eta_P$ but that is only possible if the Bethe-Salpeter
amplitude depends on it.  $\eta_P=1/2$ for charge-conjugation eigenstates.]

The general form of the Bethe-Salpeter amplitude for pseudoscalar mesons is
\begin{eqnarray}
\label{genpibsa}
\lefteqn{\Gamma_H(k;P) = }\\
&& \nonumber {\cal T}^H \gamma_5 \left[ \rule{0mm}{5mm} i E_H(k;P) +
\gamma\cdot P F_H(k;P) \right.\\
&& \nonumber \left. \rule{0mm}{5mm} + \gamma\cdot k \,k \cdot P\, G_H(k;P) +
\sigma_{\mu\nu}\,k_\mu P_\nu \,H_H(k;P) \right]\,,
\end{eqnarray}
where for bound states of constituents with equal current-quark masses, the
scalar functions $E$, $F$, $G$ and $H$ are even under $k\cdot P \to - k\cdot
P$.  [NB.\ Since the homogeneous BSE is an eigenvalue\linebreak problem,
$E_H(k;P)= E_H(k^2,k\cdot P|P^2)$, etc.; i.e., $P^2$ is not a variable,
instead it labels the solution.]

In Eq.~(\ref{genbse}), $K^{rs}_{tu}(q,k;P)$ is the renormalised,
fully-amputated quark-antiquark scattering kernel, which also appears
implicitly in Eq.~(\ref{gendse}) because it is the kernel of the
inhomogeneous DSE satisfied by $\Gamma_\nu(q;p)$.  $K^{rs}_{tu}(q,k;P)$ is a
$4$-point Schwinger function, obtained as the sum of a countable infinity of
skeleton diagrams.  It is two-particle-irreducible, with respect to the
quark-antiquark pair of lines and does not contain quark-antiquark to single
gauge-boson annihilation diagrams, such as would describe the leptonic decay
of a pseudoscalar meson.  (A connection between the fully-amputated
quark-antiquark scattering amplitude: $M = K + K ({\cal S}{\cal S}) K +
\ldots\,$, and the Wilson loop is discussed in Ref.~\cite{nora}.)

The complexity of $K^{rs}_{tu}(q,k;P)$ is one reason why quantitative studies
of the quark DSE currently employ {\it Ans\"atze} for $D_{\mu\nu}(k)$ and
$\Gamma_\nu(k,p)$.  However, as illustrated by Ref.~\cite{mrt98}, this
complexity does not prevent one from analysing aspects of QCD in a model
independent manner and proving general results that provide useful
constraints on model studies.  [NB.\ References
\cite{mrt98,pieterVM,mtpion,mtkaon,marisESI} provide a pedagogical guide to
the rigorous and practical application of BSEs to the light-quark sector of
QCD, and should be particularly helpful to practitioners for whom their
efficacy has not hitherto been apparent.]

\begin{figure}[t]
\begin{center}
\epsfig{figure=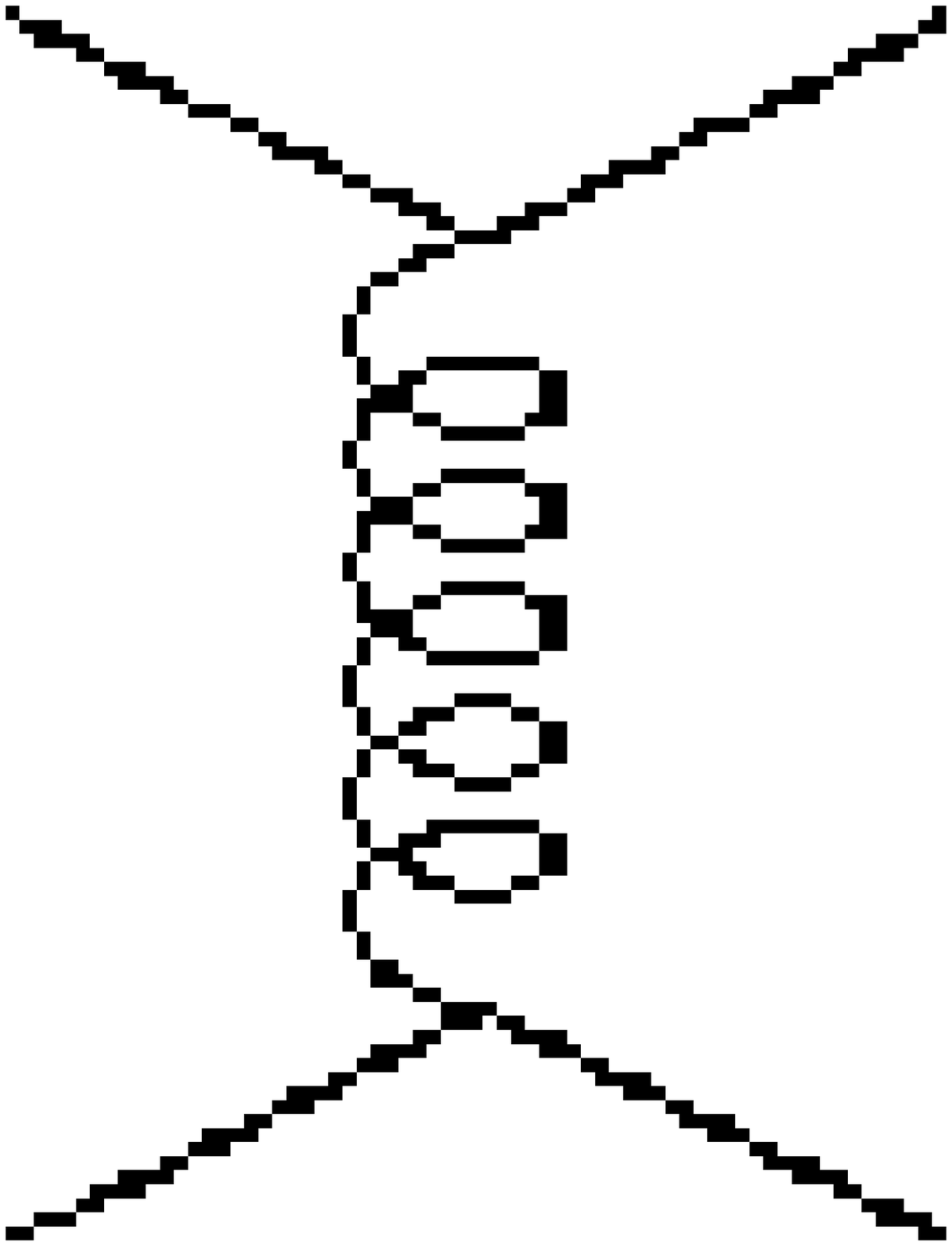,height=3.5cm}
\caption{\label{skeletona}Leading [Ladder] contribution to the systematic
expansion of the quark-antiquark scattering kernel introduced in
Ref. \protect\cite{truncscheme}.  In this expansion, the propagators are
dressed but the vertices are bare.  (Adapted from
Ref.~\protect\cite{truncscheme}.)}
\end{center}
\end{figure}

$K$ has a skeleton expansion in terms of the elementary, dressed-particle
Schwinger functions; e.g., the dressed-quark and -gluon propagators.  The
first two orders in one systematic expansion \cite{truncscheme} are depicted
in Figs.~\ref{skeletona}, \ref{skeletonb}.  This particular expansion, in
concert with its analogue for the kernel in the quark DSE, provides a means
of constructing a kernel that, order-by-order in the number of vertices,
ensures the preservation of vector and axial-vector Ward-Ta\-ka\-ha\-shi
identities.  This is particularly important in QCD where the Goldstone boson
nature of the pion must be understood as a {\it consequence} of its internal
structure, and without fine-tuning; i.e., the masslessness of the pion in the
chiral limit cannot arise as the result of carefully varying parameters in a
putative potential.  The Goldstone boson character of the pion is easily
understood via the interrelation between the pseudoscalar BSE and quark DSE
\cite{truncscheme,mr97,mrt98}.

\FIGURE[ht]{\epsfig{figure=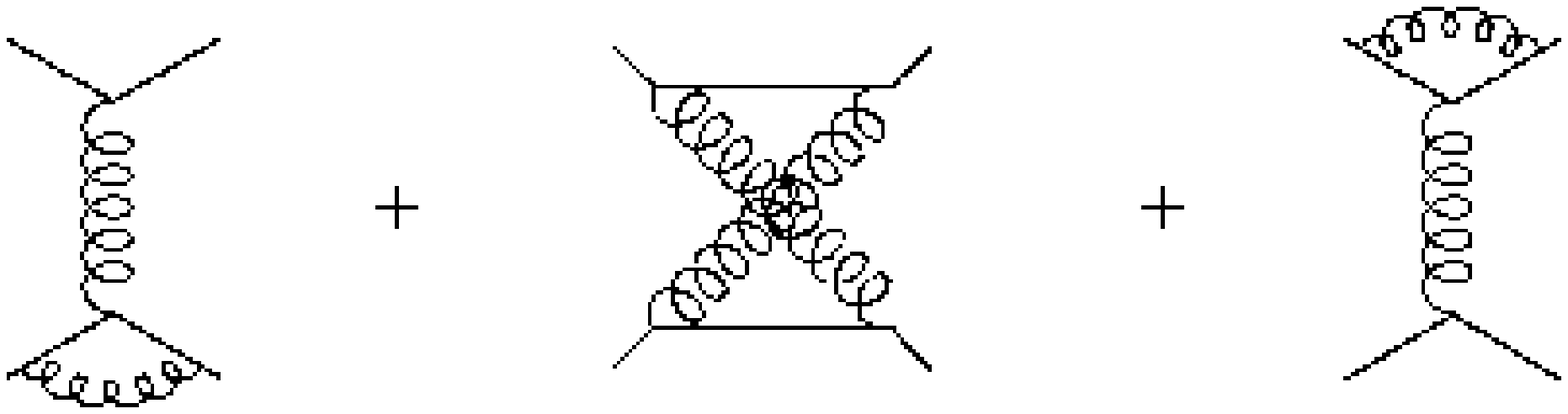,height=3.0cm}\caption{\label{skeletonb}Next-to-leading
order contribution in the truncation scheme of
Ref. \protect\cite{truncscheme}.  (Adapted from
Ref.~\protect\cite{truncscheme}.)}}

Analysing the flavour-nonsinglet inhomogeneous axial-vector BSE and its
pseudoscalar analogue one obtains, via the axial-vector Ward-Ta\-ka\-ha\-shi
identity, a mass formula, exact in QCD, for pseudoscalar mesons
\cite{mr97,mrt98,miranskymunczek}:\label{pcacstart}
\begin{equation}
\label{massform}
f_H \,m_H^2 =  {\cal M}_H^\zeta \,r^\zeta_H\,,
\end{equation}
${\cal M}_H^\zeta={\rm tr}_F [ M_\zeta \{{\cal T}^H,({\cal T}^H)^{\rm t}\}]$,
where ${\cal T}^H$ is a matrix that describes the flavour content of the
meson; e.g., ${\cal T}^{\pi^+}= \sfrac{1}{2}(\lambda_F^1+i\lambda_F^2)$,
$(\cdot)^{\rm t}$ denotes matrix transpose, and
\begin{equation}
\label{caint}
\sfrac{1}{\surd 2}\,f_H P_\mu = Z_2\,{\rm tr}\int^\Lambda_k\, \left({\cal
T}^H\right)^{\rm t} \gamma_5 \gamma_\mu \,\chi_H(k;P)\,,
\end{equation}
\begin{equation}
\label{rHres}
\begin{array}{rcl}
r^\zeta_H  & = &
-i\sqrt{2}\,Z_4\,{\rm tr}\int^\Lambda_k \left({\cal T}^H\right)^{\rm t}
\gamma_5 \, \chi_H(k;P)\\
 & & \\
 & := &  - 2i\,\langle\bar q q \rangle_\zeta^H\, \frac{1}{f_H}\,.
\end{array}
\end{equation}
Importantly, the expression in Eq. (\ref{massform}) is valid for {\it all}
values of the current-quark mass.  [NB.\ In this section the normalisation is
such that $f_\pi = 131\,$MeV.]

Equation (\ref{caint}) gives the formula for the re\-si\-due of the meson's
pole in the axial-vector vertex, which completely determines the strong
interaction contribution to its leptonic decay.  The factor of $Z_2$ on the
r.h.s.\ is just that necessary to ensure that $f_H$ is independent of the
renormalisation point, regularisation mass-scale and gauge parameter; i.e.,
to ensure that $f_H$ is truly a physical observable.  Its intuitive character
is also plain: it is the gauge-invariant projection of the meson's
Bethe-Salpeter wave function at the origin.

Equation (\ref{rHres}) is the expression for the re\-si\-due of the meson's
pole in the pseudoscalar vertex.  In this case again the factor $Z_4$ on the
r.h.s.\ depends on the gauge parameter, the regularisation mass-scale and the
renormalisation point.  This dependence is exactly that required to ensure
that: 1) $r_H$ is finite in the limit $\Lambda\to \infty$; 2) $r_H$ is
gauge-parameter independent; and 3) the renormalisation point dependence of
$r_H$ is just such as to ensure that the r.h.s.\ of Eq.\ (\ref{massform}) is
renormalisation point {\it independent}.

These formulae exemplify the manner in\linebreak which gauge invariant
results are obtained from the gauge covariant Schwinger functions the DSEs
yield.

It is straightforward to show that in the chiral limit, defined in
Eq.\ (\ref{chirallimit}), Eq.\ (\ref{rHres}) yields
\begin{equation}
r_H^\zeta \stackrel{\hat m \to 0}{\to} 
=  \frac{2}{f_\pi^0}\,\langle\bar q q \rangle^0_\zeta\,,
\end{equation}
where $\langle\bar q q \rangle^0_\zeta$ is the vacuum quark condensate, Eq.\
(\ref{qbq0}), and $f_\pi^0$ is the pseudoscalar meson decay constant in the
chiral limit.  Hence, as a corollary of Eq.\ (\ref{massform}) one recovers
\cite{mrt98} what is commonly called the ``Gell-Mann--Oakes--Renner''
relation:
\begin{equation}
\label{gmor}
f_H^2 m_H^2 = 4 \,m_\zeta\, \langle\bar q q \rangle^0_\zeta + O(\hat
m^2)\,.
\end{equation}

Another particularly important result is that the ax\-i\-al-vector
Ward-Takahashi identity also constrains the chiral limit behaviour of the
scalar functions in Eq. (\ref{genpibsa}).  This is a pointwise manifestation
of Goldstone's theorem with \cite{mrt98}:
\begin{eqnarray}
\label{bwti}
\sfrac{1}{\surd 2} f_\pi^0 E_\pi(k;0) &= & B(k^2)\,, \\
 F_R(k;0) +  \surd 2 \, f_\pi^0 F_\pi(k;0)  & = & A(k^2)\,, \\
G_R(k;0) +  \surd 2 \,f_\pi^0 G_\pi(k;0)    & = & 2 A^\prime(k^2),\\
\label{gwti} 
H_R(k;0) +  \surd 2 \,f_\pi^0 H_\pi(k;0)    & = & 0\,.
\end{eqnarray}
In these identities $(\cdot)_\pi$ denotes solution functions obtained in the
chiral limit and $(\cdot)_{R}$ are functions appearing in the pole-free part
of the inhomogeneous axial-vector vertex.

The formula in Eq.\ (\ref{bwti}) is a quark-level Gold\-ber\-ger-Treiman
relation, exact in QCD, while the next two identities indicate that
pseudoscalar me\-sons necessarily have a pseu\-do\-vec\-tor component.  This
is crucial because it is these amplitudes that dominate the mesons'
electromagnetic form factors at large meson energy and yield \cite{mrpion}
\begin{equation}
\label{piFFuv}
q^2 F_H(q^2) = \,{\rm constant~at~large-}q^2\,,
\end{equation}
up to logarithmic corrections, making a connection with the result of
perturbative QCD analyses.  If the pseudovector amplitudes are ignored one
finds instead \cite{cdrpion} that $(q^2)^2 F_H(q^2) = \,$con\-stant at
large-$q^2$.  These results are illustrated in Figs.\ \ref{piFFs},
\ref{piFFl}.
\FIGURE[ht]{\epsfig{figure=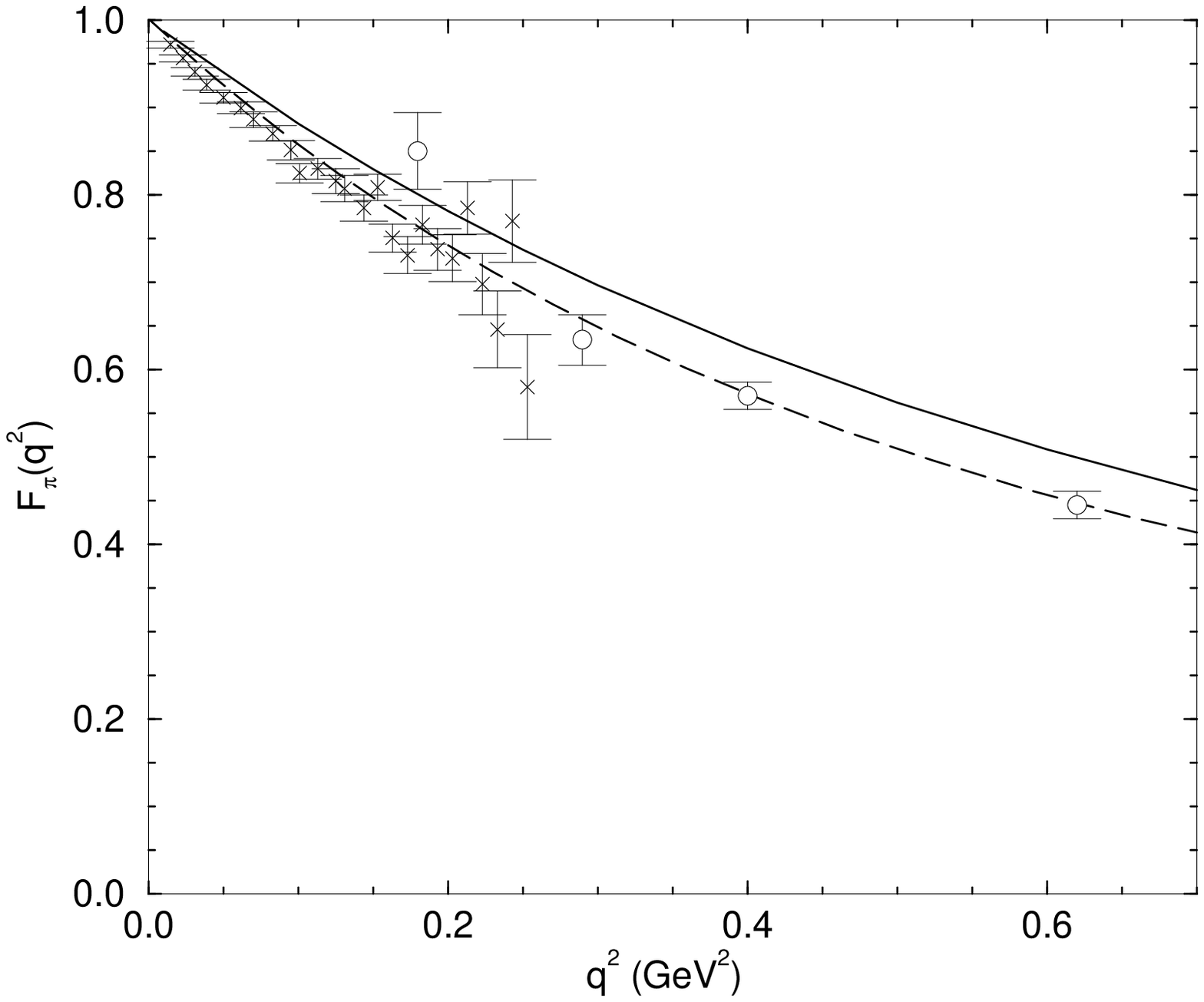,height=6.2cm}\caption{\label{piFFs}Small-$q^2$
pion form factor calculated in Ref.\ \protect\cite{mrpion} and compared with
data from Refs.\ \protect\cite{bebek} (circles) and \protect\cite{amend}
(crosses).  The solid line is the result obtained including the pseudovector
components while the dashed line was obtained without them
\protect\cite{cdrpion}.  Clearly they contribute little on this domain.
(Adapted from Ref.\ \protect\cite{mrpion}.)}}
\FIGURE[ht]{\epsfig{figure=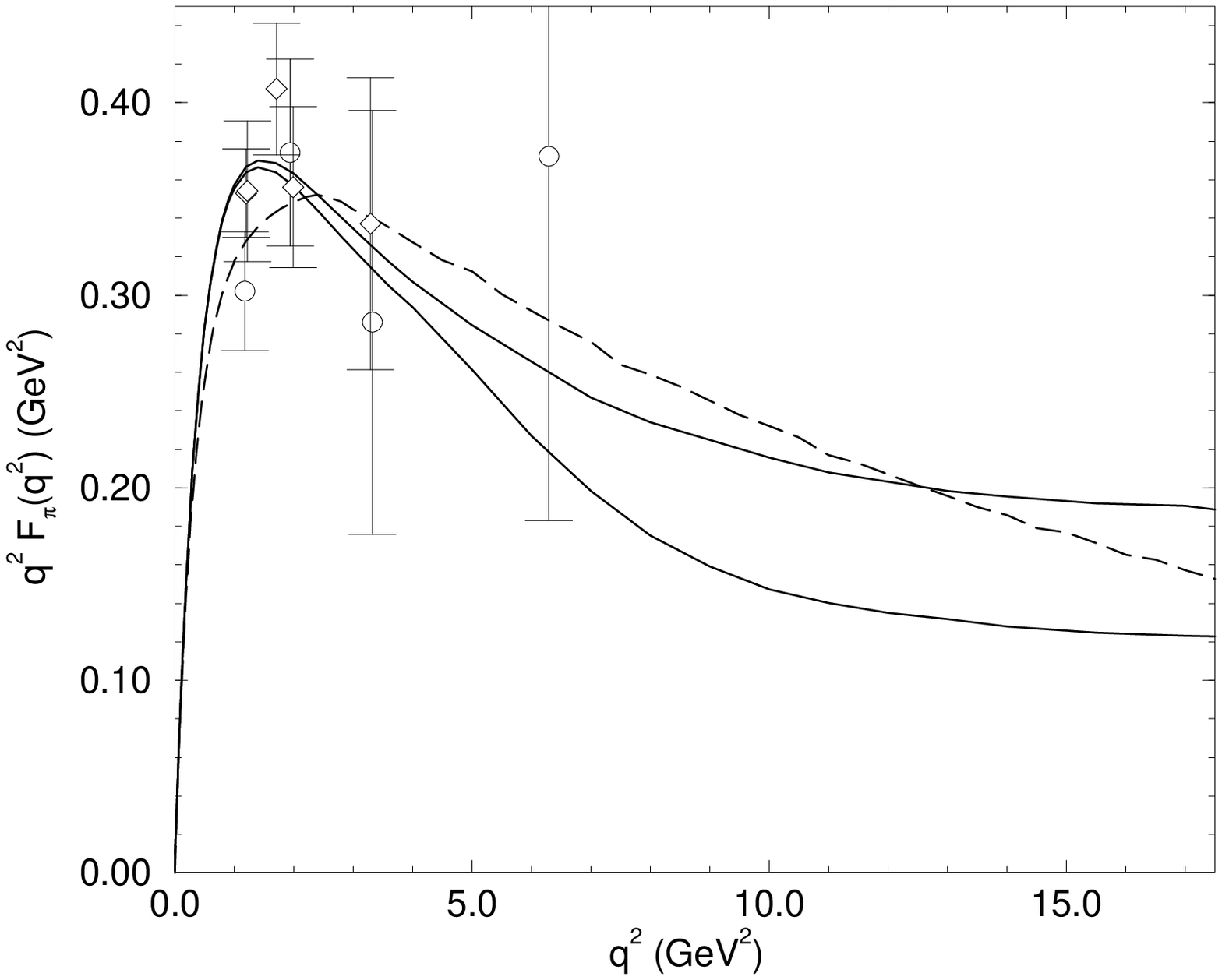,height=5.76cm}\caption{\label{piFFl}Calculated
\protect\cite{mrpion} form factor compared with the largest $q^2$ data
available: diamonds - Ref.\ \protect\cite{bebek}; and circles - Ref.\
\protect\cite{bebekB}.  The solid lines are the results obtained when the
pseudovector components of the pion Bethe-Salpeter amplitude are included
(two limiting models were employed in the calculation), the dashed-line when
they are neglected \protect\cite{cdrpion}.  On this domain the difference is
plain, with only the solid lines exhibiting Eq.\ (\protect\ref{piFFuv}).
(Adapted from Ref.\ \protect\cite{mrpion}.)}}
They are, of course, contingent upon Eq. (\ref{Mchiral}), which describes the
correct ultraviolet behaviour of the mass function in QCD.  Any model that
employs/predicts a mass function that falls faster than a single power of
$1/p^2$ will be in conflict with perturbative analyses of the form factors.
[NB.\ No bound state amplitude can fall slower than $1/p^2$.]

\subsection{Heavy-quark Limit}
\label{sub.heavy}
After this focus on the light-quark sector I now return to Eqs.\
(\ref{massform})--(\ref{rHres}) and note that they also have two important
corollaries valid in the limit of large cur\-rent-quark mass.  To elucidate
them one rewrites the total momentum
\begin{equation}
P_\mu = m_H v_\mu = (\hat M_Q + E_H)\,,
\end{equation}
where: $m_H$ is the heavy-hadron's mass; $\hat M_Q$ is a
constituent-heavy-quark mass, $\hat M_Q \approx M^E_Q\approx \hat m_Q$ for
heavy-quarks, as illustrated in Fig.\ \ref{massfunction}; and $E_H$ is the
``binding energy.''  

Now the dressed propagator for the heavy-quark in the heavy-meson can be
written \cite{usPLB}
\begin{equation}
\label{hqf}
S_Q(k+P) = \frac{1}{2}\,\frac{1 - i \gamma\cdot v}{k\cdot v - E_H}
+ {\rm O}\left(\frac{|k|}{\hat M_{Q}},\frac{E_H}{\hat M_{Q}}\right)\,,
\end{equation}
where $k$ is the momentum of the lighter constituent, and the canonically
normalised Bethe-Salpeter amplitude can be expressed as
\begin{equation}
\label{hmbsa}
\Gamma_H(k;P) = \sqrt{m_H}\; \Gamma_H^{<\infty}(k;P)
\end{equation}
where $\Gamma_H^{<\infty}(k;P)$ is pointwise-finite in the limit
$m_H\to\infty$.  It is obvious that in the calculation of observables the
meson's Bethe-Salpeter amplitude will limit the range of $|k|$ so that
Eq. (\ref{hqf}) will be a good approximation if-and-only-if both the momentum
space width of the amplitude and the binding energy are significantly
less-than $\hat M_Q$.

\FIGURE[ht]{\epsfig{figure=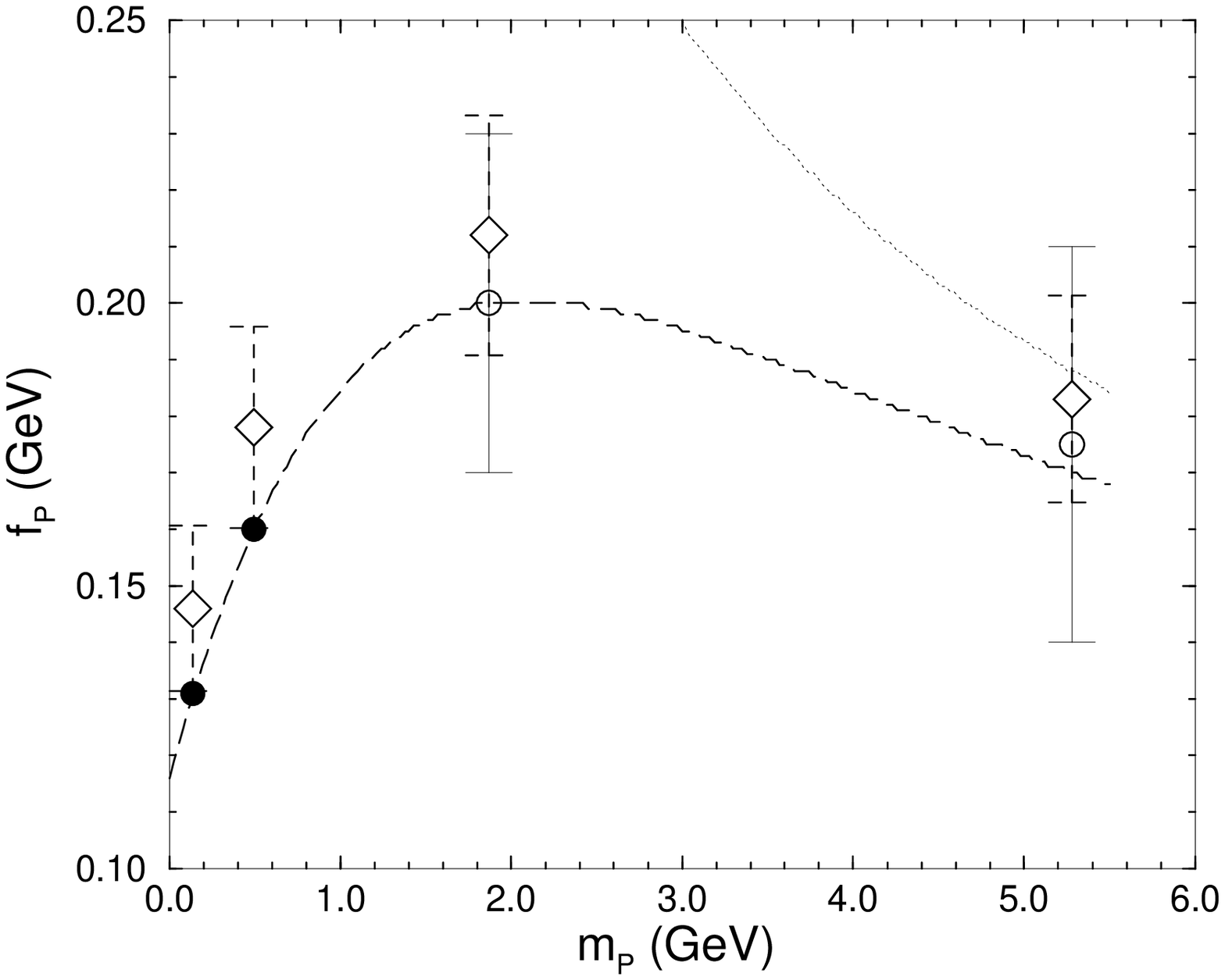,height=5.9cm}\caption{\label{figfH}Mass-dependence
of the meson leptonic decay constant.  Experimental values of $f_{\pi,K}$,
filled circles; lattice estimates \protect\cite{FS} of $f_{D}= 0.20 \pm
0.03\,$GeV and $f_B= 0.17 \pm 0.035\,$GeV, open circles; decay constants
calculated in Ref. \protect\cite{mishaSVY} with an estimated 10\% theoretical
error.  The dashed line is an eye-guiding fit to the experimental and lattice
estimates, which exhibits the large-$m_H$ limit of Eq. (\protect\ref{hqfH}).
The dotted line is the large-$m_H$ limit of this fit.  (Adapted from Ref.\
\protect\cite{mishaSVY}.)}}
Using Eqs.\ (\ref{hqf}) and (\ref{hmbsa}) in Eq.\ (\ref{caint})\linebreak
yields
\begin{equation}
\label{hqfH}
f_H = \frac{c_H^f}{\sqrt{m_H}}\,,
\end{equation}
where $c_H^f$ is a calculable and finite constant that depends only on
$\Gamma_H^{<\infty}(k;P)$ and the dressed-light-quark propagator.  In Eq.\
(\ref{hqfH}) the DSEs reproduce a well-known general consequence of
heavy-quark symmetry.  Since $f_\pi < f_K$ is an experimental fact neither
this formula nor the limits that lead to it are valid for light mesons.  The
actual situation, as determined in Ref. \cite{mishaSVY}, is depicted in
Fig. \ref{figfH}.  Clearly, finite current-quark mass corrections are
significant for the $c$-quark.  In fact, such corrections can be $\lsim 30$\%
in $b\to c$ transitions and as much as a factor of two in $c \to d$
transitions \cite{mishaSVY}.

Applying the same analysis to Eq.\ (\ref{rHres}) gives
\begin{equation}
\label{hqrH}
r_H = c_H^{r_\zeta} \, \sqrt{m_H}\,,
\end{equation}
where again $c_H^r$ is a calculable and finite constant.  Using this and
Eq. (\ref{hqfH}) in Eq. (\ref{massform}) one finds
\cite{mishaSVY,marisadelaide} that in the heavy-quark limit
\begin{equation}
\label{dsemass}
m_H = \frac{c_H^{r_\zeta}}{c_H^f} {\cal M}_H^\zeta\,.
\end{equation}
Thus the quadratic trajectory of Eq. (\ref{gmor}) evolves into a linear
trajectory when the current-quark mass becomes large; i.e., the mass of a
heavy-meson rises linearly with the mass of its heaviest constituent.  This
is illustrated in Fig. \ref{figmH}, which indicates that a linear trajectory
may describe all but the $\pi$-meson.  That was anticipated in Ref.\
\cite{mr97} and later clarified in an explicit calculation
\cite{pieterrostock} where the onset of the linear trajectory was observed at
$2$--$3$-times the $s$-quark current-mass.
\FIGURE[ht]{\epsfig{figure=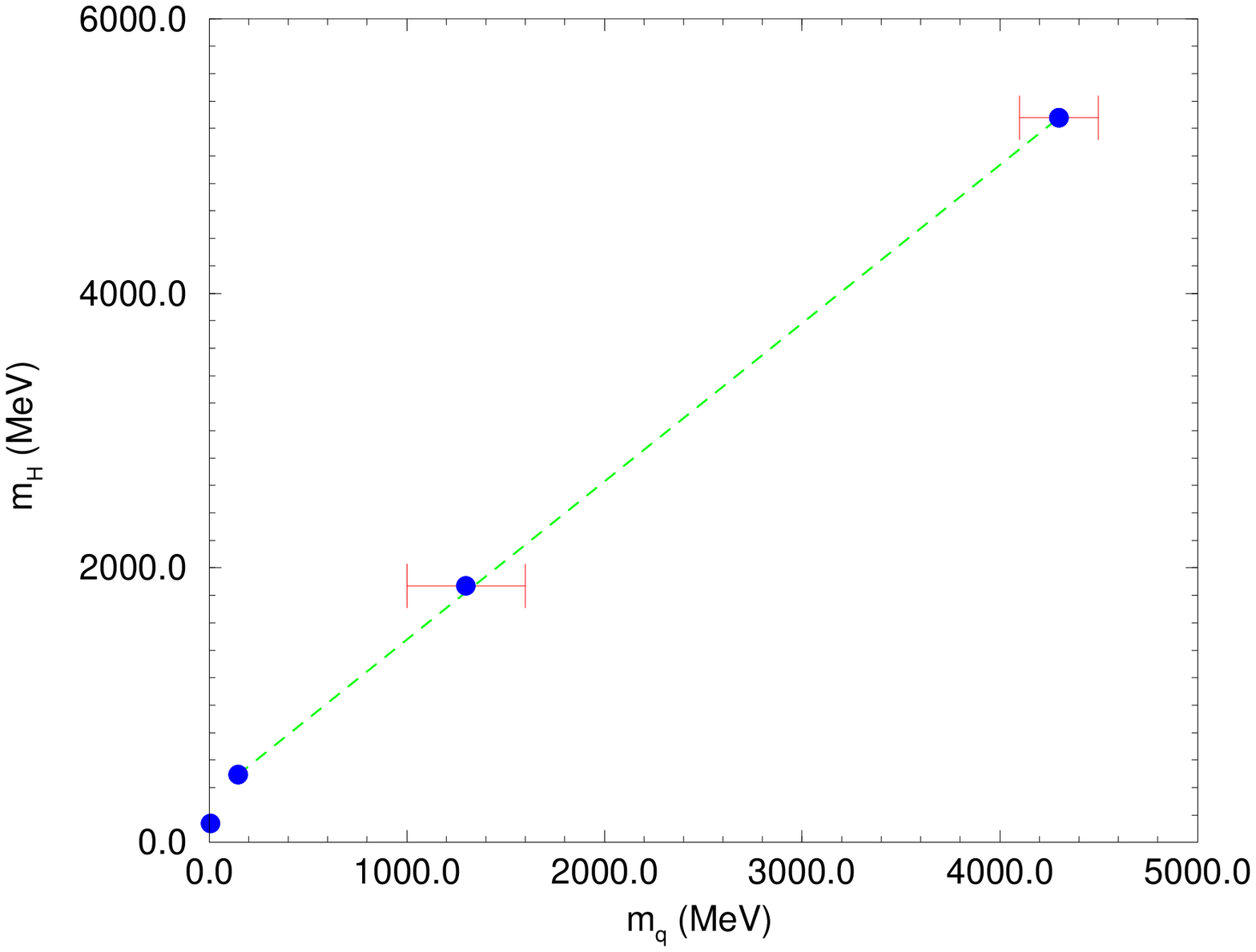,height=5.5cm}\caption{\label{figmH}Mass-dependence
of meson masses.  Experimental value of the masses and estimates of $m_q$,
are taken from Ref.\ \protect\cite{pdg98}.  The straight line, drawn through
the $D$- and $B$-meson masses, depicts the linear trajectory predicted in
Eq.\ (\protect\ref{dsemass}).  (Adapted from Ref.\
\protect\cite{marisadelaide}.)}}

Explicit calculations, of course, require that the form of the scattering
kernel be specified.  The dressed-ladder form
\begin{eqnarray}
\label{ourkernel}
\lefteqn{K_{tu}^{rs}(q,k;P)= }\\
&& \nonumber {\cal G}(k^2) \,D^{\rm free}_{\mu\nu}(p-q)\,
\left[\sfrac{1}{2}\lambda^a \gamma_\mu\right]_{tr}
\left[\sfrac{1}{2}\lambda^a \gamma_\nu\right]_{su}\,,
\end{eqnarray}
in concert with Eq.\ (\ref{ouransatz}) in the quark DSE, ensures that the
vector and axial-vector Ward-Ta\-ka\-ha\-shi identities are satisfied.
Preserving these identities is important because it guarantees current
conservation in electromagnetic processes\linebreak and an explicit
realisation of the consequences of DCSB as elucidated herein.  The choice in
Eq.\ (\ref{ourkernel}) is that yielded by the truncation scheme of Ref.\
\cite{truncscheme} and clearly maintains the ultraviolet behaviour of QCD,
Eq.\ (\ref{aqquv}).

With an explicit form for the kernel, Eq. (\ref{genbse}) is an eigenvalue
problem and it has solutions only for particular, separated values of $P^2$.
The eigenvector associated with each eigenvalue: $\Gamma_H(k;P)$, the
Bethe-Salpeter amplitude, is a one-particle-irreducible, fully-amputated
quark-\linebreak me\-son vertex.  In the flavour non-singlet pseudoscalar
channels the solutions having the lowest eigenvalues correspond to the $\pi$-
and $K$-mesons, while in the vector channels they correspond to the
$\omega$-, $\rho$- and $\phi$-mesons.

To calculate meson masses, one first solves Eqs.~(\ref{dsemod}) and
(\ref{sigmod}) for the renormalised dres\-sed-quark propagator.  This
numerical solution for $S(p)$ is then used in the BSE [obtained using
Eqs.~(\ref{genbse}), (\ref{genpibsa}), (\ref{ourkernel})], which for
pseudoscalar mesons is a coupled set of four homogeneous equations, one set
for each meson.  Solving the equations is a challenging numerical exercise,
requiring careful attention to detail, and two complementary methods were
both used in Refs.~\cite{mr97,pieterVM}.  While the numerical methods were
identical, the authors of Ref.~\cite{pieterVM} used a simplified version of
the effective interaction:
\begin{eqnarray}
\label{gk2VM}
\lefteqn{\frac{{\cal G}(k^2)}{k^2} = \frac{4\pi^2}{\omega^6} D k^2 {\rm
e}^{-k^2/\omega^2}}\\
&& \nonumber + 4\pi\,\frac{ \gamma_m \pi} {\sfrac{1}{2} \ln\left[\tau +
\left(1 + k^2/\Lambda_{\rm QCD}^2\right)^2\right]} {\cal F}(k^2) \,,
\end{eqnarray}
and varied the single parameter $D$ along with the current-quark masses:
$\hat m_u=\hat m_d$ and $\hat m_s$, in order to reproduce the observed values
of $m_{\pi}$, $m_K$ and $f_\pi$ listed on page \ref{bsefit}.  The many other
calculated results in Refs.\ \cite{pieterVM,mtpion,mtkaon} are {\it
predictions} in the sense that they are unconstrained.  A particular feature
of these studies is that the Poincar\'e covariant four-dimensional BSE is
solved directly eschewing the commonly used ar\-te\-fice of a
three-dimensional reduction, which introduces spurious effects when imposing
compatibility with\linebreak Goldstone's theorem and can also lead to the
misinterpretation of a model's parameters~\cite{alkoferbse}.

\section{Epilogue}
\label{sec.contemporary}
I have focused on some robust qualitative aspects of Dyson-Schwinger equation
[DSE] studies.  A compelling complement is Ref.\ \cite{marisESI}, which
describes in detail the direct application of the sin\-gle-pa\-ra\-me\-ter
DSE-model of Eq.\ (\ref{gk2VM}) to the spectrum and dynamical properties of
light me\-sons.  Additional important aspects are described in Ref.\
\cite{a1b1}: in particular the use of vector meson polarisation observables
to probe the long-range part of the quark-quark interaction, which highlights
the intuitive character of Bethe-Salpeter amplitudes; and also in Refs.\
\cite{mrpion,cdrpion,anomalies} where the manner in which the DSEs provide
for the intrinsic preservation of current-algebra's anomalies is elucidated
and exemplified.  Light-baryon properties too have been studied, with Ref.\
\cite{reinhard} demonstrating that a quark-diquark simplification \cite{reg}
of the covariant baryon Fadde$^\prime$ev equation is a good foundation for
spectroscopy, and Refs.~\cite{jacquesmartin} proving it efficacious for a
wide range of scattering observables.

In recent years there has also been a renewal of interest in DSE applications
to QCD at non\-zero chemical potential and temperature, which herein I have
not mentioned at all.  In closing therefore I note that, as will have become
clear from Secs.\ \ref{sec.dcsb} and \ref{sec.conf}, confinement and
dynamical chiral symmetry breaking are simultaneously accessible using the
DSEs.  This was exploited in Ref.\ \cite{axelT} wherein a well-constrained
DSE-model of QCD was used to explore the deconfinement and chiral symmetry
re\-sto\-ra\-tion transitions at $T\neq 0$: in this case the transitions are
coincident and second order, which is easily understood via the causal
relation between these two phenomena and the behaviour of the dressed-quark
mass function.

More recently the nonzero temperature and density applications have expanded
to include hadron properties, as illustrated in Refs.\ \cite{marisT}, and
have become more refined.  A large part of Ref.\ \cite{bastirev} is a
chronicle of this.  Reference \cite{bastirev} also highlights a contemporary
challenge in this area: the development of a DSE-based quantum transport
theory.  That quest is driven by a desire to incorporate the treatment of
out-of-equilibrium aspects of a relativistic heavy ion collision into a
framework that reliably describes the chemical and thermal equilibrium
features of hot and dense strongly-interacting matter.

\medskip

{\bf Acknowledgments.}~~I am grateful for the hospitality and support of the
Erwin Schr\"o\-dinger Institute for Mathematical Physics, Vienna, during my
two engagements at this workshop, and observe that it is indeed true that
kangaroos are rare in Austra$\!\!\!/$l$\!\!\!/$ia.  In preparing this
contribution I have drawn heavily on research completed in collaboration with
F.T.\ Hawes,
M.A.\ Ivanov,
Yu.L.\ Kalinovsky,
P.\ Maris,
S.M.\ Schmidt
and P.C.\ Tandy.
This work was supported by the US Department of Energy, Nuclear Physics
Division, under contract no. W-31-109-ENG-38.

\appendix
\section{Euclidean Metric Conventions}
\label{appEM} For $4$-vectors $a$, $b\,$:
\begin{equation}
a\cdot b := a_\mu\,b_\nu\,\delta_{\mu\nu} := \sum_{i=1}^4\,a_i\,b_i\,,
\end{equation}
so that a spacelike vector, $Q_\mu$, has $Q^2>0$.  The Dirac matrices are
Hermitian and defined by the algebra
\begin{equation}
\{\gamma_\mu,\gamma_\nu\} = 2\,\delta_{\mu\nu}\,;
\end{equation}
and 
\begin{equation}
\gamma_5 := -\,\gamma_1\gamma_2\gamma_3\gamma_4
\end{equation}
so that
\begin{equation}
{\rm tr}\left[ \gamma_5 \gamma_\mu\gamma_\nu\gamma_\rho\gamma_\sigma \right] 
= - 4 \,\varepsilon_{\mu\nu\rho\sigma}\,,\;
\varepsilon_{1234}= 1\,.
\end{equation}
The Dirac-like representation of these matrices is:
\begin{equation}
\vec{\gamma}=\left(
\begin{array}{cc}
0 & -i\vec{\tau}  \\
i\vec{\tau} & 0
\end{array}
\right),\;
\gamma_4=\left(
\begin{array}{cc}
\tau^0 & 0 \\
0 & -\tau^0
\end{array}
\right),
\end{equation}
where the $2\times 2$ Pauli matrices are:
\begin{equation}
\label{PauliMs}
\begin{array}{cc}
\tau^0 = \left(
\begin{array}{cc}
1 & 0 \\
0 & 1
\end{array}\right),\; &
\tau^1 = \left(
\begin{array}{cc}
0 & 1 \\
1 & 0
\end{array}\right),\; \\
\tau^2 = \left(
\begin{array}{cc}
0 & -i \\
i & 0
\end{array}\right),\; &
\tau^3 = \left(
\begin{array}{cc}
1 & 0 \\
0 & -1
\end{array}\right).
\end{array}
\end{equation}

\label{References}

\end{document}